\catcode`@=11
\long\def\@makecaption#1#2{
   \vskip 10pt
   \setbox\@tempboxa\hbox{{\small\bf #1.} \ {\small #2}}
   \ifdim \wd\@tempboxa >\hsize       
   {\small\bf #1.} \ {\small #2}\par  
   \else                              
        \hbox to\hsize{\hfil\box\@tempboxa\hfil}
   \fi}
\catcode`@=12

\catcode`@=11
\def\secteqno{\@addtoreset{equation}{section}%
\def\theequation{\thesection.\arabic{equation}}}
\def\endsecteqno{\def\theequation{\@ifundefined{chapter}%
{\arabic{equation}}{\thechapter.\arabic{equation}}}}
\newcounter{subequation}
\def\thesubequation{\alph{subequation}}
\def\sneqnarray{\stepcounter{equation}\let\@currentlabel=\theequation
\setcounter{subequation}{1}
\def\@eqnnum{{\rm (\theequation\thesubequation)}}
\global\@eqcnt\z@\tabskip\@centering\let\\=\@eqncr\let\@@eqncr=\@@sneqncr
$$\halign to \displaywidth\bgroup\@eqnsel\hskip\@centering
 $\displaystyle\tabskip\z@{##}$&\global\@eqcnt\@ne
 \hskip 2\arraycolsep \hfil${##}$\hfil
 &\global\@eqcnt\tw@ \hskip 2\arraycolsep
$\displaystyle\tabskip\z@{##}$\hfil
tabskip\@centering&\llap{##}\tabskip\z@\cr}
\def\endsneqnarray{\@@sneqncr\egroup $$\global\@ignoretrue}
\def\@@sneqncr{\let\@tempa\relax
   \ifcase\@eqcnt \def\@tempa{& & &}\or \def\@tempa{& &}
   \else \def\@tempa{&}\fi
     \@tempa \if@eqnsw\@eqnnum\stepcounter{subequation}\fi
     \global\@eqnswtrue\global\@eqcnt\z@\cr}
\def\nobiblabels{\def\@lbibitem[##1]##2{\@bibitem{##2}}}
\catcode`@=12

  \def\pa{\partial}   \def\dag{\dagger}
    
\def\bnabla{{\bm \nabla}}

\def\nn{\nonumber}

\newcommand{\cev}[1]{\reflectbox{\ensuremath{\vec{\reflectbox{\ensuremath{#1}}}}}} 

\documentclass[aps,10pt,prd,amsmath,amssymb,preprintnumbers,superscriptaddress,nofootinbib]{revtex4-2}

\usepackage[german,english]{babel}
\usepackage{hyperref}
\usepackage{bm}
\usepackage{slashed}
\usepackage{multirow}
\usepackage{epsfig}
\usepackage{bbm}
\usepackage{bbold}
\usepackage{xcolor}
\usepackage{tensor}
\usepackage{slashbox}

\def\shrinkage{2.1mu}
\def\vecsign{\mathchar"017E}
\def\dvecsign{\smash{\stackon[-1.95pt]{\mkern-\shrinkage\vecsign}{\rotatebox{180}{$\mkern-\shrinkage\vecsign$}}}}
\def\dvec#1{\def\useanchorwidth{T}\stackon[-4.2pt]{#1}{\,\dvecsign}}
\usepackage{stackengine}
\stackMath
\usepackage{graphicx}
\def\nbg{\dvec{\nabla}}

\begin{document}

\title{Heavy meson thresholds in Born-Oppenheimer effective field theory}

\author{Jaume Tarr\'us Castell\`a}
\email{jtarrus@iu.edu}
\affiliation{Department of Physics, Indiana University, Bloomington, Indiana 47401, USA}
\affiliation{Center for Exploration of Energy and Matter, Indiana University, Bloomington, Indiana 47408, USA}

\date{\today}

\begin{abstract}
We consider heavy meson-antimeson pairs and their coupling to quarkonium in the context of nonrelativistic EFTs incorporating the adiabatic expansion. We work out all the leading order couplings of quarkonium to heavy meson-antimeson pairs and obtain their contributions to the masses and widths of quarkonia. We match the new potentials terms to NRQCD. Using the available lattice data for the coupled system of quarkonium and the lowest lying heavy meson-antimeson pair, we extract the mixing potential and use it to compute numerically the contributions of $D\bar{D}(B\bar{B})$ and $D_s\bar{D}_s(B_s\bar{B}_s)$ to the masses and widths of the charmonium (bottomonium) states for $l=0,1,2$ and up to $n=6$ covering the states in threshold region. When a quarkonium state and a heavy meson-antimeson pair are separated by small energy gaps, their interactions can be described by a threshold EFT with contact interactions. We work out the matching between the two EFTs obtaining the couplings of the threshold EFT in terms of the mixing potential and quarkonium wave functions.
\end{abstract}

\maketitle

\section{Introduction}\label{sec:int}

The discovery during the past two decades of several dozen exotic hadrons, understood as those that cannot be classified as mesons or baryons in the quark model picture, has made evident an important gap in our understanding of the QCD spectrum and therefore of its underlying dynamics. Many of these states have been discovered in the double-heavy sector in experiments at B-factories (BaBar, Belle, and CLEO), $\tau$-charm facilities (CLEO-c and BESIII) and hadron colliders (CDF, D0, LHCb, ATLAS, and CMS), see Ref.~\cite{Brambilla:2019esw} for a review on the experimental status. Since the creation of heavy quarks pairs in hadrons is highly suppressed due to their mass being much larger than $\Lambda_{\rm QCD}$, the number of heavy quarks in a hadron can be identified by the hadron mass while the light-quark and gluonic content can be identified from other quantum numbers. Therefore, the identification of an exotic state is more straightforward in the heavy quark sectors. In the charmonium and bottomonium sectors these exotic states are commonly labeled as ``XYZ'' and appear in the mass region of the heavy meson-antimeson pairs also known as heavy meson thresholds. 

\begin{figure}[ht!]
\centerline{\hspace{2.5cm}\includegraphics[width=0.8\linewidth]{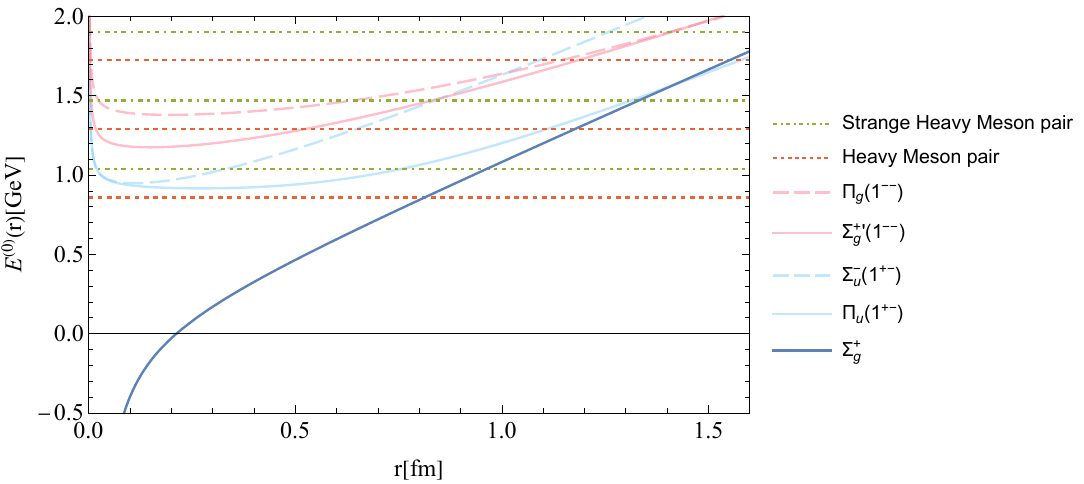}}
\caption{Sketch of the spectrum of static energies for a heavy quark-antiquark pair. The static energies corresponding to standard and hybrid quarkonium states are labeled by their $D_{\infty h}$ representation and in the latter case by the quantum numbers of the gluelump in parenthesis. The shapes are obtained from a fit to the lattice data of Ref.~\cite{Juge:2002br}. The heavy meson-antimeson static energies are drawn as constant lines at the energies given by the spin and isospin averages of the heavy meson and antimeson minus the heavy quark mass corresponding to the sum of the $\bar{\Lambda}$ parameter (defined in Eq.~\eqref{s:ths:e7}) for each heavy meson. The three energy levels for the heavy meson-antimeson pairs without and with closed strangeness correspond to the three blocks of static states of heavy meson pairs in Table~\ref{hmpreps}.
}
\label{plot_ses}
\end{figure}

A crucial tool in understanding the spectrum of double-heavy hadrons is the adiabatic expansion between the dynamics of the heavy quarks and that of the light-degrees of freedom, either light-quarks or gluons. In this picture the double-heavy hadrons are the heavy quark bound states supported by the spectrum of static energies (also known as adiabatic surfaces) associated to the light degrees of freedom. Therefore, the first step to elucidate the quarkonium spectrum in the threshold region is to determine the spectrum of static energies. These have been studied on the lattice and the emerging picture, which we sketch in Fig.~\ref{plot_ses}, is as follows. The ground state corresponds to the standard quarkonium potential, while the first excited static state corresponds to a heavy meson-antimeson pair~\cite{Bali:2005fu,Bulava:2019iut}. At higher energies additional static energies corresponding to pairs of heavier heavy mesons should also appear. The static energies of the heavy meson pairs appear as horizontal lines at the energy corresponding to the heavy meson-antimeson pair mass with some possible attractive or repulsive behavior for short distances~\cite{Bali:2005fu}. Beyond the quarkonium sector, the static energies of heavy meson-antimeson pairs have also been studied in the lattice for $I=1$ in Refs.~\cite{Bali:2005fu,Prelovsek:2019ywc,Sadl:2021bme} and for heavy meson-meson in Refs.~\cite{Bicudo:2012qt,Brown:2012tm,Bicudo:2015vta,Bicudo:2015kna,Bicudo:2021qxj}. Excited heavy-quark-antiquark static energies, corresponding to hybrid quarkonium states also appear in the region above the first heavy meson threshold~\cite{Bali:2000vr,Juge:2002br,Capitani:2018rox,Schlosser:2021wnr}. These are repulsive in the short-distance due to the heavy quarks being in a color octet state but in the long-distance become a linear, confining potential, corresponding to the string excitations of the standard quarkonium potential. The spectrum of hybrid quarkonium static energies is formed by multiplets of static energies corresponding to different gluonic states, refereed as gluelumps~\cite{Foster:1998wu}.

The standard and hybrid quarkonium spectra above the first heavy meson threshold contains enough states to account for all the observed neutral heavy exotic states observed so far~\cite{Oncala:2017hop}, however the specific assignation of states to the observed states is not yet clear. One important step to clarify the exotic spectrum is to incorporate the mixing between the different static states. The mixing of standard quarkonium with the lowest lying hybrid static energies was considered in Ref.~\cite{Oncala:2017hop}. Due to the $1^{+-}$ quantum numbers of the associated gluelump, the mixing is through a heavy-quark spin dependent operator which is $1/m_Q$ suppressed, with $m_Q$ being the heavy quark mass. However, this is not the case for the next set of hybrid static energies, associated to the $1^{--}$ gluelump, which can mix at leading order with standard quarkonium. 

We will focus on the mixing of standard quarkonium with heavy meson pairs. As we will discuss, standard quarkonium couples at leading order with the lowest lying heavy meson pairs. Therefore, this coupling is the most important effect of the threshold region degrees of freedom on the quarkonium masses up to the energies where $1^{--}$ gluelump hybrid states start to be relevant.
This problem has been studied in Refs.~\cite{Bicudo:2019ymo,Bicudo:2020qhp,Bicudo:2022ihz,Bruschini:2020voj,Bruschini:2021sjh} using models based on an extension of the Born-Oppenheimer approximation that includes the mixing potentials between the heavy meson pair and quarkonium, which is also refereed as the diabatic approach. In Refs.~\cite{Bicudo:2019ymo,Bicudo:2020qhp,Bicudo:2022ihz} the potentials where extracted from the lattice data of Ref.~\cite{Bali:2005fu} and the coupled channel scattering problem was solved numerically. From the poles of the heavy meson $t$ matrices for specific angular momenta, the bottomonium spectrum was identified. In the case of in Refs.~\cite{Bruschini:2020voj,Bruschini:2021sjh}, a model for the quarkonium heavy meson pair mixing was used. To obtain the spectra a mix approach was used in which the contributions to a particular quarkonium state of the above-lying thresholds are obtained solving the coupled channel Schrödinger equations while the below-lying ones are obtained in perturbation theory.

In this paper we examine the coupling of quarkonium to the heavy meson pairs in the context of an effective field theory (EFT) incorporating the heavy quark mass and adiabatic expansions. The EFT incorporating these two expansions for quarkonium is known as strongly coupled potential nonrelativistic QCD (pNRQCD)~\cite{Brambilla:2000gk,Pineda:2000sz} while its extension to nontrivial light degrees of freedom content has been called Born-Oppenheimer EFT (BOEFT)~\cite{Brambilla:2017uyf,Soto:2020pfa,Soto:2020xpm}. The content of this paper can be considered an extension of either EFT, however for easy reference we will consider it as part of the latter. At leading order in the heavy-quark mass expansion the heavy mesons are characterized by the spin, parity and flavor of the light-quark state~\cite{Neubert:2000hd}. Combining these for a heavy meson-antimeson pair one arrives to the total spin, parity and charge conjugation of the light-quarks which characterize the heavy meson-antimeson state. We derive all the leading order couplings of heavy-meson-antimeson states to quarkonium. Using these, we compute the contribution of the heavy meson thresholds to the quarkonium self-energy from which we obtain the contributions to the quarkonium masses and decay widths. We obtain the matching expression of the mixing potential in NRQCD considering both the cases when the mixing can be considered a perturbation and not. We compute numerically the quarkonium spectrum up to ${\cal O}(1/m_Q)$ for $S,\,P$ and $D$ waves and up to the principal quantum number $n=6$ which covers the mass range for which exotic quarkonium states have been discovered. To do so, we use a parametrization of the potentials that combines lattice data and fix order computations. For these states we compute the contributions of the lowest lying heavy meson pairs without and with closed strangeness, using the mixing potential from the lattice data of Ref.~\cite{Bali:2005fu}. Finally, we work out the matching of BOEFT with a threshold EFT containing a quarkonium state, a heavy meson and a heavy antimeson as effective degrees of freedom with contact interactions, as used for instance in Refs.~\cite{Fleming:2008yn,Guo:2009wr,Guo:2010ak,Mehen:2011yh,Mehen:2011tp,Mehen:2013mva,Guo:2013zbw,Mehen:2015efa,Chen:2016mjn}. We discuss what expansion is involved and under what conditions it can be implemented.

We organize the paper as follows. In Sec.~\ref{sec:ths} we show how to incorporate heavy meson pairs into BOEFT and compute their contribution to the masses and widths of quarkonium states  in perturbation theory. The matching of the new potentials to NRQCD is discussed in Sec.~\ref{sec:mnrq}. In Sec.~\ref{sec:num} we extract the form of the static potentials for standard quarkonium and the mixing potential with the lowest lying heavy meson pairs from lattice QCD together with inputs from perturbation theory. Using these potentials, in Sec.~\ref{sec:nr} we compute the contributions of the lowest lying heavy meson thresholds to the quarkonium masses and decay widths. The matching of BOEFT to the threshold EFT is examined in Sec.~\ref{s:teft}. Finally, we provide our conclusions on Sec.~\ref{sec:con}.

\section{Heavy meson pairs in BOEFT}\label{sec:ths}

To construct the Lagrangian for BOEFT one first identifies the quantum numbers of the light degrees of freedom, chiefly the spin $\kappa$, parity $p$ and charge conjugation $c$, that generate the set of static energies we are interested in. For the case of heavy meson-antimeson pairs this means identifying the light-quark states. From the heavy meson heavy-quark-spin multiplets we can identify the quantum numbers of the light-quark states. The ground state corresponds to $\kappa^p=(1/2)^+$ and is followed by two states of similar mass with $\kappa^p=(1/2)^-$ and $\kappa^p=(3/2)^-$. Each heavy meson-antimeson pair is characterized by the spin and parity of the light-quark states forming the two heavy mesons, which we label as $\kappa_1^{p_1}$ and $\kappa_2^{p_2}$. Combining the spin and parity of the light quark and antiquark we arrive to the allowed total spin, parity and charge conjugation, $\kappa^{pc}$, of the light-quarks in Table~\ref{hmpreps}. Each of these combinations is represented as a field in BOEFT. We will denote these fields as ${\cal M}_{\kappa^{pc}}(t,\,\bm{r},\,\bm{R})$, with $\bm{r}$ and $\bm{R}$ being the relative coordinate and center of mass of the heavy quarks. One should keep in mind that the fields ${\cal M}_{\kappa^{pc}}$ have spin indices corresponding to the light-quarks (in the $\kappa$ representation) and corresponding to the heavy-quarks (in the $(1/2)^{*}\otimes (1/2)$ representation). The field ${\cal M}_{\kappa^{pc}}$ carries the light-quark flavor quantum numbers of the heavy meson-antimeson it corresponds, this can be the individual light-quark flavors, isospin or chiral symmetry representations. Note that in the latter cases, the field ${\cal M}_{\kappa^{pc}}$ would correspond to a sum of heavy meson-antimeson pairs. The light-quark flavor quantum numbers do not affect the construction of the Lagrangian, and to simplify the notation we will not track them in this section. However, it should be keep in mind that the parameters and potentials of the heavy meson-antimeson pair field do depend on the light-quark flavor. The quarkonium field  will be denoted as $\Psi(t,\,\bm{r},\,\bm{R})$. The bilinear terms for both fields can be read off Ref.~\cite{Soto:2020xpm}, with quarkonium corresponding to the $\kappa^{pc}=0^{++}$ case:
\begin{align}
{\cal L}=&\Psi^{\dagger}\left[i\partial_t-h_{\Psi}\right]\Psi+\sum_{\kappa^{pc}}{\cal M}^{\dagger}_{\kappa^{pc}}\left[i\partial_t-h_{\kappa^{pc}}\right]{\cal M}_{\kappa^{pc}}\,,\label{boeft}
\end{align}
The expansion of the Hamiltonian densities $h_{\Psi}$ and $h_{\kappa^p}$ up to $1/m_Q$ is as follows
\begin{align}
h_{x}=\frac{\bm{p}^2}{m_Q}+\frac{\bm{P}^2}{4m_Q}+V_{x}^{(0)}(\bm{r})+\frac{1}{m_Q}V_{x}^{(1)}(\bm{r},\,\bm{p})\,,\quad x=\Psi,\,\kappa^{pc}\label{hamden}
\end{align}
with $\bm{p}=-i\bnabla_r$ and $\bm{P}=-i\bnabla_R$.

\begin{table}[ht!]
\begin{tabular}{c|c|c}  \hline\hline
$\kappa_1^{p_1}\otimes\kappa_2^{p_2}$ & $\kappa^{pc}$ & $D_{\infty h}$ \\ \hline
$(1/2)^+\otimes(1/2)^+$         & $0^{-+}$ & $\Sigma_u^-$ \\
                                & $1^{--}$ & $\Sigma_g^+,\,\Pi_g$ \\ \hline
$(1/2)^+\otimes(1/2)^-$         & $0^{++}$ & $\Sigma_g^+$\\                                 
                                & $1^{+-}$ & $\Sigma_u^-,\,\Pi_u$ \\
$(1/2)^+\otimes(3/2)^-$         & $1^{+-}$ & $\Sigma_u^-,\,\Pi_u$  \\
                                & $2^{++}$ & $\Sigma_g^+,\,\Pi_g,\,\Delta_g$\\ \hline
$(1/2)^-\otimes(3/2)^-$         & $1^{--}$ & $\Sigma_g^+,\,\Pi_g$  \\
                                & $2^{-+}$ & $\Sigma_u^-,\,\Pi_u,\,\Delta_u$ \\ 
$(1/2)^-\otimes(1/2)^-$         & $0^{-+}$ & $\Sigma_u^-$ \\
                                & $1^{--}$ & $\Sigma_g^+,\,\Pi_g$ \\
$(3/2)^-\otimes(3/2)^-$         & $0^{-+}$ & $\Sigma_u^-$ \\
                                & $1^{--}$ & $\Sigma_g^+,\,\Pi_g$ \\
                                & $2^{-+}$ & $\Sigma_u^-,\,\Pi_u,\,\Delta_u$ \\
                                & $3^{--}$ & $\Sigma_g^+,\,\Pi_g,\,\Delta_g,\,\Phi_g$\\ \hline\hline
\end{tabular}
\caption{Total spin, parity, charge conjugation, and $D_{\infty h}$ representations of the light quark-antiquark pair combinations of the three lightest light-quark states forming heavy mesons. Each block of states, separated by a single horizontal line, corresponds to degenerate or nearly degenerate states.}
\label{hmpreps}
\end{table}

The symmetry group of two static heavy quarks is $D_{\infty h}$, which is a cylindrical symmetry group for rotations along the $\hat{\bm{r}}$ axis. The representations of $D_{\infty h}$ are labeled as $\Lambda^{\sigma}_\eta$, where $\Lambda$ is the absolute value of the projection into the heavy quark-antiquark axis of the spin $\kappa$ of the light degrees of freedom and is labeled by capital Greek letters: $\Sigma$, $\Pi$, $\Delta,...$ corresponding to $\Lambda=0,1,2,...$ . $\eta$ is the CP eigenvalue, denoted by $g = + 1$ and $u = - 1$. Finally, for $\Lambda=0$, there is a reflection symmetry with  respect to a plane passing through the $\hat{\bm{r}}$ axis. The eigenvalues of the corresponding symmetry operator being labeled as $\sigma=\pm 1$. The potential terms in Eq.~\eqref{hamden} should be expanded in representations of $D_{\infty h}$. For quarkonium there is only one possible projection, into the $\Sigma_g^+$ representation, and therefore the corresponding projector is just an identity. Furthermore, both the leading order and next-to-leading order potentials are heavy-quark spin independent and therefore consists of a single potential term.

For the heavy meson pairs all the possible projections for each $\kappa^{pc}$ state are listed in the third column in Table~\ref{hmpreps}. The static potential between a heavy meson-antimeson pair has been studied on the lattice in Refs.~\cite{Bali:2005fu,Bicudo:2012qt,Brown:2012tm,Bicudo:2015vta,Bicudo:2015kna,Prelovsek:2019ywc,Bicudo:2021qxj,Sadl:2021bme}. The results show that the potentials are mostly flat lines at the energy corresponding to the heavy meson masses except in some cases in the short-distance limit where the potential is attractive or repulsive depending on the specific heavy meson pair. The hidden heavy flavor isospin $I=0$ case has been studied in Refs.~\cite{Bali:2005fu,Bulava:2019iut}. As we will discuss in detail in Sec.~\ref{qssp}, once the mixing with quarkonium is taken into account, the heavy meson-antimeson static potential is completely flat for the range of data in Ref.~\cite{Bali:2005fu}. Therefore, for this work we will assume that the interaction between the heavy mesons is negligible. Thus, we set the heavy meson-antimeson static potential to be the sum of the heavy meson masses at leading order in in the heavy quark mass expansion minus the origin of energies, which is set at the heavy quark masses
\begin{align}
V_{\kappa^{pc}}^{(0)}(\bm{r})=\left(\bar{\Lambda}_{\kappa^{p_1}_1}+\bar{\Lambda}_{\kappa^{p_2}_2}\right)\mathbb{1}_\kappa\,,\label{s:ths:e6}
\end{align}
with $\mathbb{1}_\kappa$ an identity in the light-quark spin-space, and $\bar{\Lambda}_{\kappa^p}$ is related to the heavy meson masses~\cite{Neubert:2000hd} as
\begin{align}
m_{H_{\kappa^p}}=m_Q+\bar{\Lambda}_{\kappa^p}+{\cal O}(1/m_Q)\,.\label{s:ths:e7}
\end{align}
Notice that, due to this identity all the projections into $D_{\infty h}$ representations of the field ${\cal M}_{\kappa^{pc}}$ have degenerate static potentials. We assume that the subleading potential $V_{\kappa^{pc}}^{(1)}$ corresponds to the sum of the spin-dependent ${\cal O}(1/m_Q)$ operators in the Hamiltonian of each heavy meson corresponding to ${\cal M}_{\kappa^{pc}}$. This is equivalent to assume that there is no significant heavy meson-antimeson interaction at this order either. These spin-dependent operators are the ones that break the degeneracy between different total spin heavy meson states.

The results of Table~\ref{hmpreps} are valid for any light-quark flavor content, however, the $\bar{\Lambda}$ value does depend on the light-quark flavor of the heavy mesons and therefore so does the position of the corresponding static energy on the spectrum of static energies.

Now let us discuss the mixing terms between quarkonium and the heavy meson-antimeson pair. Since the quarkonium field $\Psi$ inherently belongs to a $\Lambda=0$ representation, we should project the heavy meson pair field into the same representation. This can be achieved with the projection vector $P_{\kappa 0}$~\cite{Brambilla:2017uyf,Soto:2020xpm}, which is defined by the eigenvalue equation
\begin{align}     
\left(\hat{\bm{r}}\cdot\bm{S}_\kappa\right) P_{\kappa \lambda}=\lambda P_{\kappa \lambda }\,,\quad \lambda=-\kappa,...,\kappa\,,
\end{align}
with $\Lambda=|\lambda|$ and $\bm{S}_\kappa$ the spin operator for the $\kappa$ representation. From textbooks, as for instance Ref.~\cite{Varshalovich:1988}, one can find that
\begin{align}     
(P_{\kappa 0})_\alpha=i^\kappa\sqrt{\frac{4\pi}{2\kappa+1}}Y_{\kappa\alpha}(\hat{\bm{r}})\,,\label{sec:ths:e1}
\end{align}
where $Y_{\kappa\alpha}(\hat{\bm{r}})$ is a spherical harmonic. Notice, that one can think of $Y_{\kappa\alpha}(\hat{\bm{r}})$ as rank $\kappa$ irreducible tensor made out of powers of $\hat{\bm{r}}$, with $\alpha$ acting as the spin index. For instance, for $\kappa=1$ the projection vector is just $(P_{1 0})_\alpha=i\hat{\bm{r}}_\alpha$. The phase in Eq.~\eqref{sec:ths:e1} is arbitrary and is chosen so the projection vector transforms under time reversal as a spin-$\kappa$ irreducible tensor, see Appendix~\ref{a1}.

The most general leading order operator containing $\Psi$ and ${\cal M}_{\kappa^{pc}}$ which is invariant under $D_{\infty h}$ and $O(3)$ transformations and discrete symmetries is as follows
\begin{align}
{\cal L}_{\rm mix}=&-\int d^3\bm{r}\sum_\kappa V_{\rm mix}^{\kappa}(r)\left\{({\cal M}^\dag_{\kappa^{\pm\pm}})^{\alpha} (P_{\kappa 0})_\alpha\Psi+\Psi^\dag (P^*_{\kappa 0})^\alpha({\cal M}_{\kappa^{\pm\pm}})_{\alpha}\right\}\,.\label{sec:ths:e2}
\end{align}
Note that summation over repeated spin indices is implicit throughout the paper. More details on the notation for the spin indices can be found in Appendix~\ref{a1}. If we look at Table~\ref{hmpreps}, we can see that there is at least one coupling of each heavy meson-antimeson pair to quarkonium through the operators in Eq.~\eqref{sec:ths:e2}.

One can check that the Lagrangian in Eq.~\eqref{sec:ths:e2} is the most general set of leading order mixing operators with the following argument. The only objects one could add to the mixing terms in Eq.~\eqref{sec:ths:e2} that do not add a heavy-quark mass suppression are scalar matrices build out of $\hat{\bm{r}}$ and $\bm{S}_\kappa$. As was shown in Ref.~\cite{Soto:2020xpm} the projectors 
\begin{align}
({\cal P}_{\kappa\Lambda})\indices{_\alpha^{\alpha'}}=\sum_{\lambda=\pm\Lambda}(P_{\kappa \lambda})_\alpha(P^*_{\kappa \lambda})^{\alpha'}\,,\label{s:ths:e3}
\end{align}
form a basis for these matrices, since
\begin{align}
\left(\hat{\bm{r}}\cdot\bm{S}_{\kappa}\right)^{2n}&=\sum_{\Lambda}\Lambda^{2n}{\cal P}_{\kappa\Lambda}\,.
\end{align}
As the projection vectors $P_{\kappa \lambda}$ are orthogonal, the form of Eq.~\eqref{sec:ths:e2} is not altered by adding the projectors ${\cal P}_{\kappa\Lambda}$.

We can compare our operator for the mixing of quarkonium to the lowest lying heavy meson-antimeson pair, the $\kappa^{pc}=1^{--}$ term in Eq.~\eqref{sec:ths:e2}, with the one in Refs.~\cite{Bicudo:2019ymo,Bicudo:2020qhp,Bicudo:2022ihz,Bruschini:2020voj,Bruschini:2021sjh}. In the case of Refs.~\cite{Bicudo:2019ymo,Bicudo:2020qhp,Bicudo:2022ihz} the operator coincides except for an $i$ factor needed for invariance under timer reversal. However, the results for the masses and widths should not be affected by this phase. In the case of Refs.~\cite{Bruschini:2020voj,Bruschini:2021sjh} the mixing operator does not seem to take into account the light-quark spin state that couples with quarkonium.

The contribution of the heavy meson-antimeson pair on the quarkonium masses and widths can be computed in standard perturbation theory. First, let us define the following states:
\begin{align}
|n,\,l\rangle&=\int d^3\bm{r}d^3\bm{R}\,\psi_{n l}(\bm{r})\Psi^{\dag}(\bm{r},\,\bm{R})|0\rangle\,,\label{mw:e1}
\end{align}
with $\psi_{n l}$ the wave function solution of the Schrödinger equation
\begin{align}
\left(-\frac{\bnabla^2_r}{m_Q}+V_{\Psi}^{(0)}(\bm{r})\right)\psi_{nl}(\bm{r})&=E^{(0)}_{nl}\psi_{nl}(\bm{r})\,,\label{s:ths:e5}\\
\psi_{nl}(\bm{r})&=\phi_{nl}(r)Y_{lm_l}(\hat{\bm{r}})\label{s:ths:e4}\,,
\end{align}
where $n$ is the principal quantum number and $l(l+1)$ is the eigenvalue of $\bm{L}_{Q\bar{Q}}^2$.

For the heavy meson-antimeson pair the wave functions are plane waves labeled by the relative momentum $\bm{k}$. The partial wave decomposition of the plane wave is as follows:
\begin{align}
e^{-i\bm{k}\cdot\bm{r}}=\sum_{l}4\pi i^{-l}j_l(kr)Y_{lm_l}(\hat{\bm{r}})Y^*_{lm_l}(\hat{\bm{k}})\,,\label{mtt:e2}
\end{align}
where $j_l$ is a spherical Bessel function. Ignoring the heavy-quark spin, the total angular momentum of the heavy meson pair is $\bm{L}=\bm{L}_{Q\bar{Q}}+\bm{S}_\kappa$ and the eigenvalue of $\bm{L}^2$ is
$\ell(\ell+1)$. A heavy meson pair state with momentum $k=|\bm{k}|$ and $\ell$ total angular momentum is given by
\begin{align}
|k,\,\ell,\,\kappa\rangle&=\int d^3\bm{r}d^3\bm{R}\sum^{\ell+\kappa}_{l=|\ell-\kappa|}4\pi i^{-l}j_l(kr){\cal C}^{\ell m_\ell}_{l m_l\,\kappa -\alpha}(-1)^{\kappa-\alpha} Y_{lm_l}(\hat{\bm{r}}){\cal M}^{\dag \alpha}_{\kappa^{pc}}(\bm{r},\,\bm{R})|0\rangle\,,\label{mw:e2}
\end{align}
with ${\cal C}$ a Clebsch-Gordan coefficient. Recall that all repeated spin indices are summed. In order for the state in Eq.~\eqref{mw:e2} to have definite parity the sum over $l$ should be understood to run only over even or odd values.

Let us compute the expected value of the mixing term in the Lagrangian in Eq.~\eqref{sec:ths:e2} for the states in Eqs.~\eqref{mw:e1} and \eqref{mw:e2}.
\begin{align}
\langle n,\,l|\int d^3\bm{r}\,\Psi^\dag V^\kappa_{\rm mix}(P^*_{\kappa0})^\alpha{\cal M}_{\kappa\alpha}|k,\,\ell,\,\kappa\rangle=4\pi \delta_{l\ell }\delta_{m_lm_\ell }\sum^{l+\kappa}_{l'=|l-\kappa|}i^{(\kappa-l')}a_{nl}^{\kappa l'}(k)\,,
\end{align}
with
\begin{align}
a_{nl}^{\kappa l'}(k) \equiv {\cal C}^{l 0}_{l'0\,\kappa 0}\sqrt{\frac{(2l'+1)}{(2l+1)}}\int dr r^2\phi_{nl}(r) V^{\kappa}_{\rm mix}(r) j_{l'}(kr)\,.\label{mw:e6}
\end{align}

Now, we compute the self-energy contribution
\begin{align}
\int dtd^3\bm{R}\,e^{iEt}\langle 0|T\left\{\Psi(t,\,\bm{r},\,\bm{R})\Psi^\dag(0,\,\bm{r}',\,0)\right\}|0 \rangle=\sum_{nl}\left\{\frac{i\psi_{nl}(\bm{r})\psi_{nl}^*(\bm{r}')}{E-E_{nl}}+
\frac{i\psi_{nl}(\bm{r})}{E-E_{nl}}i{\cal A}_{nl}\frac{i\psi_{nl}^*(\bm{r}')}{E-E_{nl}}+\dots\right\}
\end{align}
where
\begin{align}
&i{\cal A}_{nl}=-i\sum^{l+\kappa}_{l'=|l-\kappa|}\frac{4\mu}{\pi}\int dk k^2\frac{(a_{nl}^{\kappa l'}(k))^2}{k^2_d-k^2}\,.\label{mw:e3}
\end{align}
with $k^2_d=2\mu (E_n+2m_Q-m_T)$ and $\mu$ and $m_T$ being the heavy meson-antimeson pair reduced and total masses, respectively. The contribution to the quarkonium state mass corresponds to the real part of Eq.~\eqref{mw:e3}
\begin{align}
E^{\kappa l'}_{nl}=\frac{4\mu}{\pi}{\cal P}\int dk k^2\frac{(a_{nl}^{\kappa l'}(k))^2}{k^2_d-k^2}\,,\label{mw:e4}
\end{align}
where ${\cal P}$ stands for Cauchy principal value. The contribution to the width of the quarkonium state is obtained from the imaginary part of Eq.~\eqref{mw:e3}. We obtain
\begin{align}
\Gamma^{\kappa l'}_{nl}= 4 \mu k_d\left(a_{nl}^{\kappa l'}(k_d)\right)^2\,.\label{mw:e5}
\end{align}
We note that a similar result has been obtained in Ref.~\cite{Bruschini:2021sjh}\footnote{The expression of the width in Eq.~(26) of Ref.~\cite{Bruschini:2021sjh} seems to be missing a $\mu$ factor.}.

\section{Matching to NRQCD}\label{sec:mnrq}

In this section we obtain $V_{\kappa^{pc}}^{(0)}$ and $V_{\rm mix}^{\kappa}$ as NRQCD~\cite{Caswell:1985ui,Bodwin:1994jh,Manohar:1997qy} correlators. For a more self-contained discussion we also reproduce the result for the quarkonium static potential which was obtained originally in Refs.~\cite{Wilson:1974sk,Fischler:1977yf,Brown:1979ya}. The matching of the quarkonium $1/m_Q$ suppressed potential can be found in Ref.~\cite{Brambilla:2000gk}.

Let us define the following NRQCD operators 
\begin{align}
\mathcal{O}_\Psi(t,\,\bm{r},\,\bm{R})=&\,\chi^{\dag}(t,\,\bm{x}_2)\phi(t,\,\bm{x}_2,\,\bm{x}_1)\psi(t,\,\bm{x}_1)\,,\label{sec:mnrq:e1}\\
\mathcal{O}_{\kappa^{pc}\alpha}(t,\,\bm{r},\,\bm{R})=&\,{\cal C}^{\kappa\alpha}_{\kappa_2\alpha_2\,\kappa_1-\alpha_1}(-1)^{\kappa_1+\alpha_1}\left[\chi^\dag(t,\,\bm{x}_2)({\cal Q}_{\kappa_2^{p_2}}(t,\,\bm{x_2}))_{\alpha_2}\right]\left[(\overline{\cal Q}^\dag_{\kappa_1^{p_1}}(t,\,\bm{x_1}))^{\alpha_1}\psi(t,\,\bm{x}_1)\right]\,,\label{sec:mnrq:e2}
\end{align}
with $\psi$ a Pauli spinor field that annihilates a heavy quark and $\chi$ the one that creates a heavy antiquark. The operators ${\cal Q}_{\kappa^{p}}$ contain the light-quark fields. For $\kappa^p=(1/2)^+,\,(1/2)^-$, and $(3/2)^-$ light-quark states, suitable operators are as follows:
\begin{align}
{\cal Q}_{(1/2)^+\alpha}(t,\bm{x})&= [P_+q(t,\bm{x})]_\alpha\,,\label{s:mnrq:e1}\\
{\cal Q}_{(1/2)^-\alpha}(t,\bm{x})&= [P_+\gamma^5q(t,\bm{x})]_\alpha\,,\label{s:mnrq:e2}\\
{\cal Q}_{(3/2)^-\alpha}(t,\bm{x})&= {\cal C}^{3/2\,\alpha}_{1\,m\,1/2\,\beta}\left[\left(\bm{e}^{\dagger}_{m}\cdot\bm{D}\right)\left(P_+q(t,\bm{x})\right)_\beta\right]\,,\label{s:mnrq:e3}
\end{align}
with $q(t,\bm{x})$ a light-quark Dirac field. The $\overline{\cal Q}_{\kappa^{p}}$ operators are obtained replacing $P_+$ by $P_-$ in Eqs.~\eqref{s:mnrq:e1}-\eqref{s:mnrq:e3}. Details on the construction of irreducible tensor products, such as the one in Eq.~\eqref{sec:mnrq:e2}, can be found in Appendix~\ref{a1}. The Wilson line $\phi$ is defined as 
\begin{align}
\phi(t,\bm{x},\bm{y})=P\left\{e^{ig\int_0^1 ds\left(\bm{x}-\bm{y}\right)\cdot \bm{A}(t,\bm{y}+s(\bm{x}-\bm{y}))}\right\}\,,
\end{align}
where $P$ is the path-ordering operator.

The operators in Eqs.~\eqref{sec:mnrq:e1} and \eqref{sec:mnrq:e2} interpolate for the quarkonium and heavy meson pair fields, respectively. The matching condition from NRQCD to BOEFT reads as
\begin{align}
\mathcal{O}_\Psi(t,\,\bm{r},\,\bm{R})&\cong\sqrt{Z_\Psi}\Psi(t,\,\bm{r},\,\bm{R})\,,\label{QQtopsi}\\
\mathcal{O}_{\kappa^{pc}\alpha}(t,\,\bm{r},\,\bm{R})&\cong\sqrt{Z_{\kappa^{pc}}}{\cal M}_{\kappa^{pc}\alpha}(t,\,\bm{r},\,\bm{R})\,.\label{mmtoM}
\end{align}
The normalization factors are in general functions of $Z= Z(\bm{r},\bm{p})$. The light-quark flavor quantum numbers must match in both sides of Eq.~\eqref{mmtoM}, therefore ${\cal M}_{\kappa^{pc}}$ corresponds to a single heavy meson-antimeson pair. For ${\cal M}_{\kappa^{pc}}$ fields belonging to isospin or chiral symmetry representations one should just consider the appropriate sums over the light-quark flavor in the left-hand side of Eq.~\eqref{mmtoM}.

Since quarkonium and heavy meson-antimeson pairs mix at leading order one could argue that these states are not the appropriate ones to describe the system and that one should work with a basis of states that diagonalize the Hamiltonian at leading order. However, we know from experience that quarkonium and heavy meson-antimeson pairs are useful states to describe the heavy quark-antiquark spectrum. Therefore, there must be some regime of $r$ in which the mixing can be treated as a perturbation. Thus, let us first assume that the strings in Eqs.~\eqref{sec:mnrq:e1} and \eqref{sec:mnrq:e2} overlap with well-separated NRQCD static eigenstates. Let us match the NRQCD and BOEFT correlators:
\begin{align}
&\langle 0|T\{\mathcal{O}_\Psi(t/2,\,\bm{r},\,\bm{R})\mathcal{O}_\Psi^{\dag}(-t/2,\,\bm{r}',\,\bm{R}')\}|0\rangle=\sqrt{Z_\Psi}\langle 0|T\{\Psi(t/2,\,\bm{r},\,\bm{R})\Psi^{\dag}(-t/2,\,\bm{r}',\,\bm{R}')\}|0\rangle \sqrt{Z_\Psi^{\dag}}\,,\label{corr1}\\
&\langle 0|T\{\mathcal{O}_{\kappa^{pc}\alpha}(t/2,\,\bm{r},\,\bm{R})\mathcal{O}^{\dag \alpha'}_{\kappa^{pc}}(-t/2,\,\bm{r}',\,\bm{R}')\}|0\rangle=\sqrt{Z_{\kappa^{pc}}}\langle 0|T\{{\cal M}_{\kappa^{pc}\alpha}(t/2,\,\bm{r},\,\bm{R}){\cal M}^{\dag\alpha'}_{\kappa^{pc}}(-t/2,\,\bm{r}',\,\bm{R}')\}|0\rangle \sqrt{Z^{\dag}_{\kappa^{pc}}}\,,\label{corr2}\\
&\langle 0|T\{\mathcal{O}_{\kappa^{pc}\alpha}(t/2,\,\bm{r},\,\bm{R})\mathcal{O}_\Psi^{\dag}(-t/2,\,\bm{r}',\,\bm{R}')\}|0\rangle=\sqrt{Z_{\kappa^{pc}}}\langle 0|T\{{\cal M}_{\kappa^{pc}\alpha}(t/2,\,\bm{r},\,\bm{R})\Psi^\dag(-t/2,\,\bm{r}',\,\bm{R}')\}|0\rangle \sqrt{Z_\Psi^{\dag}}\,.\label{corr3}
\end{align}

\begin{figure}[ht!]
\centerline{\includegraphics[width=0.8\linewidth]{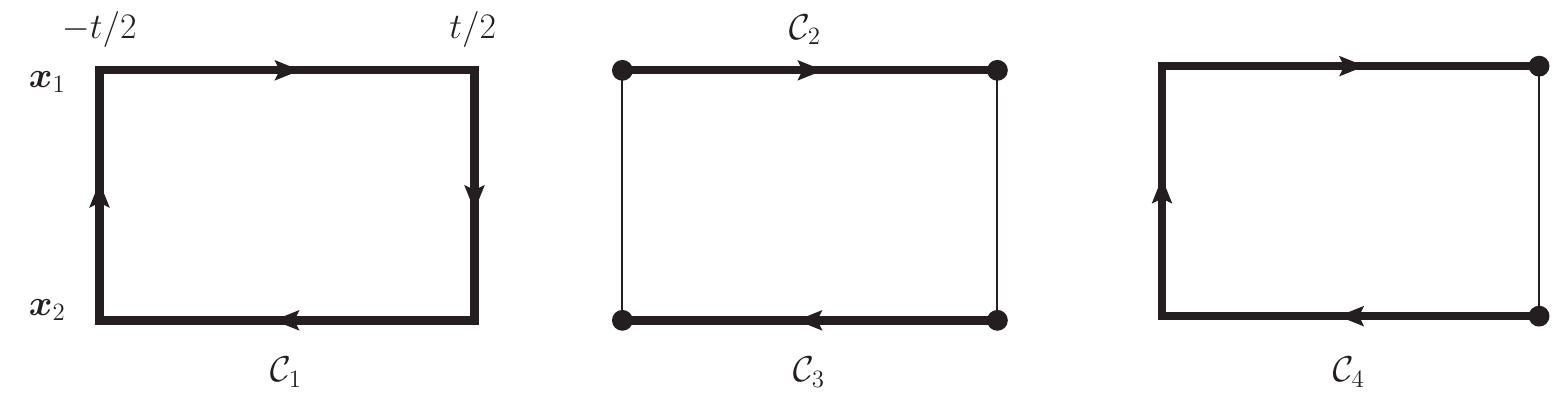}}
\caption{Wilson line paths appearing in Eqs.~\eqref{wl1}-\eqref{wl3}. The bold line represent the paths while the black dots stand for the light-quark operators.}
\label{paths}
\end{figure}

We contract the heavy-quark fields in the correlators of the right-hand side of Eqs.~\eqref{corr1}-\eqref{corr3} and define the following objects
\begin{align}
W_\Box=\,&\langle \phi_{{\cal C}_1}\rangle\,,\label{wl1}\\
\left(W^{\kappa}_=\right)\indices{_{\alpha}^{\alpha'}}=\,&{\cal C}^{\kappa\alpha}_{\kappa_2\alpha_2\,\kappa_1-\alpha_1}(-1)^{\kappa_1+\alpha_1}{\cal C}^{\kappa\alpha'}_{\kappa_2\alpha_2'\,\kappa_1-\alpha_1'}(-1)^{\kappa_1+\alpha_1'}\nn\\
&\langle (\overline{\cal Q}^\dag_{\kappa_1^{p_1}}(t/2,\,\bm{x_1}))^{\alpha_1}\phi_{{\cal C}_2}(\overline{\cal Q}_{\kappa_1^{p_1}}(-t/2,\,\bm{x_2}))_{\alpha_1'}({\cal Q}^{\dag}_{\kappa_2^{p_2}}(-t/2,\,\bm{x_1}))^{\alpha_2'}\phi_{{\cal C}_3}({\cal Q}_{\kappa_2^{p_2}}(t/2,\,\bm{x_1}))_{\alpha_2}\rangle\,,\label{wl2}\\
\left(W^{\kappa}_\sqsubset\right)_{\alpha}=\,&{\cal C}^{\kappa\alpha}_{\kappa_2\alpha_2\,\kappa_1-\alpha_1}(-1)^{\kappa_1+\alpha_1}\langle (\overline{\cal Q}^{\dag}_{\kappa_1^{p_1}}(t/2,\,\bm{x}_1))^{\alpha_1}\phi_{{\cal C}_4}({\cal Q}_{\kappa_2^{p_2}}(t/2,\,\bm{x}_2))_{\alpha_2}\rangle\,,\label{wl3}
\end{align}
with $\phi_{\cal C}$ being a Wilson line along the path ${\cal C}$
\begin{align}
\phi_{\cal C}=P\left\{e^{-ig\int_{{\cal C}}dz^{\mu}A^{\mu}(z)}\right\}\,,
\end{align}
with the paths ${\cal C}_i$, $i=1,..,4$ defined in Fig.~\ref{paths}. The right-hand sides of Eqs.~\eqref{wl1}-\eqref{wl3} are the traces of a product of color matrices and Eq.~\eqref{wl1} is a static Wilson loop.

The right-hand side of Eqs.~\eqref{corr1}-\eqref{corr3} is computed from the BOEFT Lagrangian in Eqs.~\eqref{boeft} and \eqref{hamden} and together with Eqs.~\eqref{wl1}-\eqref{wl3} we arrive at
\begin{align}
V^{(0)}_\Psi& =\lim_{t\to\infty}\frac{i}{t}\ln\left(W_\Box\right)\,,\\
V^{(0)}_{\kappa^{pc}\Lambda}& =\lim_{t\to\infty}\frac{i}{t}\ln\left({\rm Tr}\left[{\cal P}_{\kappa\Lambda}W^{\kappa}_=\right]\right)\,,
\end{align}
where the trace acts on the light-quark spin space and the projectors ${\cal P}_{\kappa\Lambda}$ are defined in Eq.~\eqref{s:ths:e3}. The mixing potential reads as
\begin{align}
V^\kappa_{\rm mix}(r)=\lim_{t\to\infty}\frac{i}{t}\frac{1}{\sqrt{W_\Box{\rm Tr}[{\cal P}_{\kappa 0}W^{\kappa}_=]}}\frac{\ln\left(\sqrt{W_\Box/{\rm Tr}[{\cal P}_{\kappa 0}W^{\kappa}_=]}\right)}{\sinh\left(\ln\sqrt{W_\Box/{\rm Tr}[{\cal P}_{\kappa 0}W^{\kappa}_=]}\right)}(P^*_{\kappa 0})^\alpha(W^{\kappa}_\sqsubset)_\alpha\,.
\end{align}

Now we consider the case when the strings in Eqs.~\eqref{sec:mnrq:e1} and \eqref{sec:mnrq:e2} overlap with two static states of similar energy
\begin{align}
\mathcal{O}_{\Psi}(\bm{r},\bm{R})|0\rangle=&\sum_{i=1,2}a^{\Psi}_i(\bm{x}_1,\bm{x}_2)|\underline{i},\Sigma_g^+;\bm{x}_1,\bm{x}_2\rangle^{(0)}\,,\label{st1m}\\
(P^*_{\kappa 0})^\alpha\mathcal{O}_{\kappa^{pc}\alpha}(\bm{r},\bm{R})|0\rangle=&\sum_{i=1,2}a^{\kappa^{pc}}_i(\bm{x}_1,\bm{x}_2)|\underline{i},\Sigma_g^+;\bm{x}_1,\bm{x}_2\rangle^{(0)}\,,\label{st2m}
\end{align}
with $|\underline{i},\Sigma_g^+;\bm{x}_1,\bm{x}_2\rangle^{(0)}$ being eigenstates\footnote{See Ref.~\cite{Brambilla:2000gk} for detailed definitions.} of the NRQCD Hamiltonian in the static limit, $H^{(0)}$, with energies $E^{(0)}_{i\Sigma_g^+}$
\begin{align}
H^{(0)}|\underline{i},\Sigma_g^+;\bm{x}_1,\bm{x}_2\rangle=E^{(0)}_{i\Sigma_g^+}(r)|\underline{i},\Sigma_g^+;\bm{x}_1,\bm{x}_2\rangle\,.\label{sec:mnrq:e4}
\end{align}
The coefficients $a^{x}_i(\bm{x}_1,\bm{x}_2)$, are the overlaps of the string $x=\Psi,\,\kappa^{pc}$ with the static state $i$.

The BOEFT potentials for $\Psi$ and $(P^*_{\kappa 0})^\alpha{\cal M}_{\kappa^{pc}\alpha}$ can be arranged as matrix
\begin{align}
\bm{V}=\left(\begin{array}{cc}
V^{(0)}_{\Psi} & V^\kappa_{\rm mix} \\
V^\kappa_{\rm mix} & V^{(0)}_{\kappa^{pc}0} \\
\end{array}\right)\,.\label{s:mnrq:e4}
\end{align}
with $V^{(0)}_{\kappa^{pc}0}={\rm Tr}[{\cal P}_{\kappa 0}V^{(0)}_{\kappa^{pc}}]$. The matrix 
\begin{align}
\bm{R}=\left(\begin{array}{cc}
\cos\theta & -\sin\theta \\
\sin\theta & \cos\theta \\
\end{array}\right)\,.
\end{align}
diagonalizes the BOEFT potential matrix in terms of the mixing angle $\theta=\theta(r)$
\begin{align}
\bm{R}^{\dag}\bm{V}\bm{R}=\left(\begin{array}{cc} V_{1\Sigma_g^+} & 0 \\ 0 &  V_{2\Sigma_g^+} \end{array}\right)\,.\label{sec:mnrq:e3}
\end{align}
The potentials from Eqs.~\eqref{s:mnrq:e4} and \eqref{sec:mnrq:e3} are related as follows
\begin{align}
V^{(0)}_{\Psi}&=\cos^2\theta\,V_{1\Sigma_g^+}+\sin^2\theta\,V_{2\Sigma_g^+}\,,\label{s:match:e1}\\
V^{(0)}_{\kappa^{pc}}&=\sin^2\theta\, V_{1\Sigma_g^+}+\cos^2\theta \,V_{2\Sigma_g^+}\,,\label{s:match:e2}\\
V^\kappa_{\rm mix}&=\sin\theta\cos\theta\left(V_{1\Sigma_g^+}-V_{2\Sigma_g^+}\right)\,.\label{s:match:e3}
\end{align}

Now we set the following matching condition
\begin{align}
|\underline{1},\Sigma_g^+\rangle&\cong \left(\cos\theta~\Psi+\sin\theta~(P^*_{l0})^\alpha{\cal M}_{\kappa^{pc}\alpha}\right)|0\rangle\,,\label{2smatch1}\\
|\underline{2},\Sigma_g^+\rangle&\cong \left(-\sin\theta~\Psi+\cos\theta~(P^*_{l0})^\alpha{\cal M}_{\kappa^{pc}\alpha}\right)|0\rangle\,,\label{2smatch2}
\end{align}
i.e., we match the NRQCD static eigenstates to the BOEFT eigenstates resulting from diagonalizing the potential matrix in Eq.~\eqref{sec:mnrq:e3}. From the matching condition in Eqs.~\eqref{2smatch1} and \eqref{2smatch2} follows that
\begin{align}
E^{(0)}_{1\Sigma_g^+} = V_{1\Sigma_g^+}\,,\quad E^{(0)}_{2\Sigma_g^+}=V_{2\Sigma_g^+}\,.\label{sec:mnrq:e7}
\end{align}

To obtain the NRQCD overlap factors $a^x_i$ in terms of BOEFT quantities we bracket Eqs.~\eqref{st1m} and \eqref{st2m} with the static states on both sides of the equations. Then the remaining bracket in the left-hand side is obtained using Eqs.~\eqref{2smatch1} and \eqref{2smatch2} and Eqs.~\eqref{QQtopsi} and \eqref{mmtoM}. We find
\begin{align}
&a^\Psi_1=\sqrt{Z_\Psi}\cos\theta\,,\quad a^\Psi_2=-\sqrt{Z_\Psi}\sin\theta\,,\label{sec:mnrq:e5}\\
&a^{\kappa^{pc}}_1=\sqrt{Z_{\kappa^{pc}}}\sin\theta\,,\quad a^{\kappa^{pc}}_2=\sqrt{Z_{\kappa^{pc}}}\cos\theta\,.\label{sec:mnrq:e6}
\end{align}

Inserting the expansion into static eigenstates in Eqs.~\eqref{st1m} and \eqref{st2m} into the correlators in left-hand side of Eqs.~\eqref{corr1} and \eqref{corr3} and using Eqs.~\eqref{sec:mnrq:e7}-\eqref{sec:mnrq:e6}, we arrive at the following expressions
\begin{align}
W_\Box&=Z_{\Psi}\left(\cos^2\theta\, e^{-it V_{1\Sigma_g^+}}+\sin^2\theta\, e^{-itV_{2\Sigma_g^+}}\right)\,,\label{s:mnrq:e8}\\
{\rm Tr}\left[{\cal P}_{\kappa 0}W^{\kappa}_=\right]&=Z_{\kappa^{pc}}\left(\sin^2\theta\, e^{-itV_{1\Sigma_g^+}}+\cos^2\theta\, e^{-itV_{2\Sigma_g^+}}\right)\\
 (P^*_{l0})^\alpha(W^{l}_\sqsubset)_\alpha&=\sqrt{Z_{\Psi}Z_{\kappa^{pc}}}\sin\theta\cos\theta\left(e^{-itV_{1\Sigma_g^+}}-e^{-itV_{2\Sigma_g^+}}\right)\,.\label{s:mnrq:e9}
\end{align}
If, using a nonperturbative technique, the left-hand side of Eqs.~\eqref{s:mnrq:e8}-\eqref{s:mnrq:e9} is computed, then one can use the parametrization of the right-hand side to fit the data and obtain $V_{1\Sigma_g^+}$, $V_{2\Sigma_g^+}$ and $\theta$ as has been done in Ref.~\cite{Bali:2005fu}.

Finally, inverting Eqs.~\eqref{s:match:e1}-\eqref{s:match:e3} we find 
\begin{align}
\theta&=\frac{1}{2}\arctan\frac{2V^l_{\rm mix}}{V^{(0)}_\Psi-V^{(0)}_{\kappa^{pc}}}\,,\label{s:mnrq:e10}\\
V_{1\Sigma_g^+}&=\frac{1}{2}\left(V^{(0)}_\Psi+V^{(0)}_{\kappa^{pc}}-\sqrt{4(V^l_{\rm mix})^2+(V^{(0)}_{\kappa^{pc}}-V^{(0)}_\Psi)^2}\right)\,,\\
V_{2\Sigma_g^+}&=\frac{1}{2}\left(V^{(0)}_\Psi+V^{(0)}_{\kappa^{pc}}+\sqrt{4(V^l_{\rm mix})^2+(V^{(0)}_{\kappa^{pc}}-V^{(0)}_\Psi)^2}\right)\,.\label{s:mnrq:e11}
\end{align}
If we expand Eqs.~\eqref{s:mnrq:e10}-\eqref{s:mnrq:e11} for $2V^l_{\rm mix}\ll |V^{(0)}_{\kappa^{pc}}-V^{(0)}_\Psi|$ and use the result in Eqs.~\eqref{s:mnrq:e8}-\eqref{s:mnrq:e9} we recover the matching expression of the first part of this section in Eqs.~\eqref{s:match:e1}-\eqref{s:match:e3} for the case of well separated static energies. Therefore the mixing can be considered as a perturbation for mixing angles close to $0$ or $\pi/2$, corresponding to the condition $2V^l_{\rm mix}\ll |V^{(0)}_{\kappa^{pc}}-V^{(0)}_\Psi|$.

\section{Computation of the quarkonium spectra}~\label{sec:num}

In this section we compute the charmonium and bottomonium spectra and wave functions and use these results to compute the contribution of the lowest lying heavy meson-antimeson pair, corresponding to the $\kappa^{pc}=1^{--}$ state in Table~\ref{hmpreps}, without and with closed strangeness to the quarkonium states masses and widths. For these two heavy meson-antimeson pairs there is available lattice data for the mixing potential from Ref.~\cite{Bali:2005fu}. To improve the overall accuracy in the determination of the quarkonium spectra in the threshold region and since the heavy meson threshold contributions are sensitive to the energy gap between a quarkonium state and the threshold we compute it up to ${\cal O}(1/m_Q)$ accuracy.

\subsection{RS' scheme}

The quarkonium spectrum at leading order is obtained by solving the Schrödinger equation in Eq.~\eqref{s:ths:e5}, for which a heavy-quark (pole) mass, $m_Q$, value must be specified. The second input for the Schrödinger equation is the static potential, $V_\Psi^{(0)}$. Both these objects suffer from renormalon ambiguities when computed in perturbation theory~\cite{Beneke:1998ui}. The total energy of a quarkonium system is a physical observable and therefore must be free of these ambiguities. At leading order the total energy is given by $E=2m_Q+V_\Psi^{(0)}$, hence the renormalon ambiguities of the heavy quark mass and the static potential cancel each other. Therefore, it is convenient to work in a scheme in which the renormalons are subtracted from these two quantities. We use the modified renormalon subtraction scheme (RS') of Ref.~\cite{Pineda:2001zq}. The subtracted heavy-quark mass and the static potential are defined as follows:
\begin{align}
m_Q=&m^{\rm RS'}_Q(\nu_f)+\delta m^{\rm RS'}_Q(\nu_f)\,,\\
V_{\Psi({\rm p.t})}^{(0)}=&V_{\rm RS'}^{(0)}(\nu_f)-2\delta m^{\rm RS'}_Q(\nu_f)\,.
\end{align}
All the quantities must be computed to the same order in $\alpha_s$ and at the same renormalon subtraction scale $\nu_f$. We use the expressions up to ${\cal O}(\alpha_s^4)$ that can be found in Ref.~\cite{Pineda:2013lta}. We work with $\nu_f=0.7$~GeV and take the heavy quark mass values $m^{\rm RS'}_c=1.592(41)$~GeV and $m^{\rm RS'}_b=4.949(41)$~GeV determined in Ref.~\cite{Peset:2018ria} and the normalization of the renormalon  $N_m=0.5626(260)$ from Ref.~\cite{Ayala:2014yxa}. The values of $\alpha_s$ in the $\overline{\rm MS}$ scheme are obtained using \texttt{RunDec} at $4$-loop accuracy~\cite{Chetyrkin:2000yt,Herren:2017osy}.

\subsection{Static potentials}\label{qssp}

\begin{figure}[ht!]
\centerline{\includegraphics[width=0.6\linewidth]{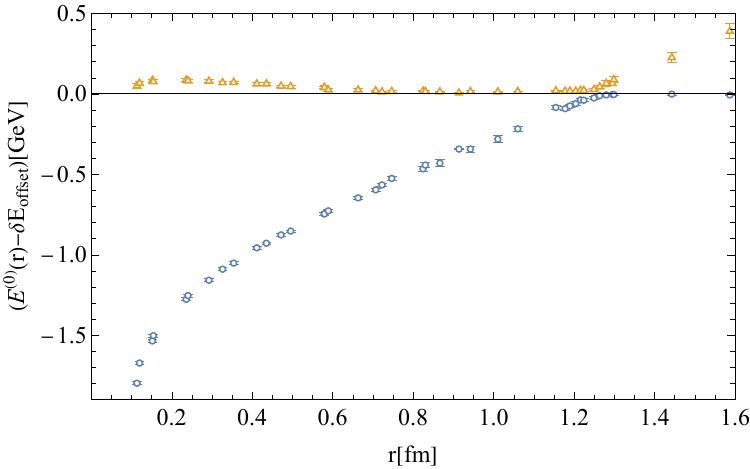}\includegraphics[width=0.4\linewidth]{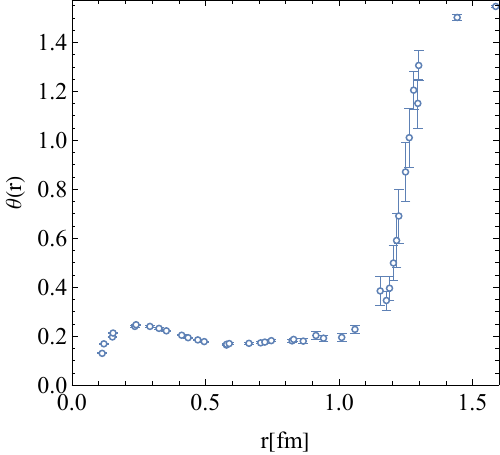}}
\caption{Plot of the lattice data of Ref.~\cite{Bali:2005fu}. In the left-hand side we plot the data for the ground and first excited static energies in the quarkonium sector. The open blue circles and the open yellow triangles correspond to the ground and first excited states respectively. Note that, in Ref.~\cite{Bali:2005fu} the origin of energies is set at the energy of the heavy meson pair for the largest $r$ computed. In the right-hand side we plot the data for the mixing angle.}
\label{lat_data}
\end{figure}

\begin{figure}[ht!]
\centerline{\includegraphics[width=0.7\linewidth]{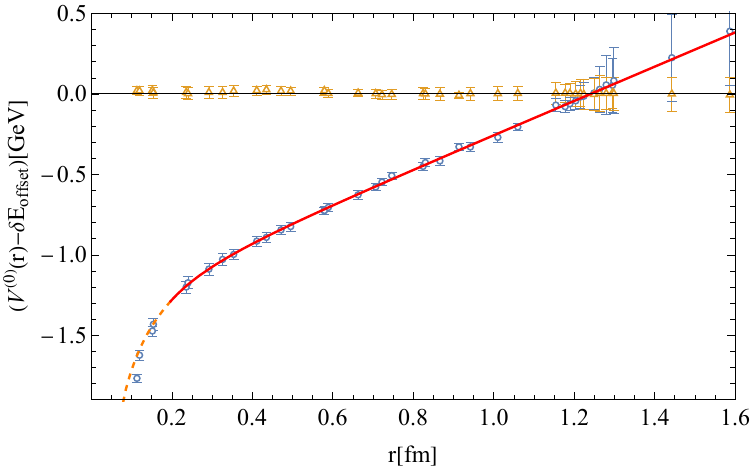}}
\caption{Plot of the static potentials. The blue open circles and the yellow triangles correspond to the lattice data of Ref.~\cite{Bali:2005fu} transformed into the quarkonium and heavy meson static potentials using Eqs.~\eqref{s:match:e1} and \eqref{s:match:e2}, respectively. The lines correspond to our parametrization of the quarkonium static potential. The dashed orange line is the perturbative potential, shifted by $\delta E_{\rm offset}$ to match the scale of the lattice data, plotted up to the matching point $r_m$. The continuous red line is the expression in Eq.~\eqref{lat_pot} fitted to the lattice data plotted from the matching point onward. The full potential formed by the orange and red lines corresponds to Eq.~\eqref{full_pot}. Note that, in Ref.~\cite{Bali:2005fu} the origin of energies is set at the energy of the heavy meson pair for the largest $r$ computed.}
\label{QQbpotplot}
\end{figure}

In Ref.~\cite{Bali:2005fu} the static energies of the heavy quark-antiquark pair coupled to a heavy meson-antimeson pair were studied using lattice QCD. The lattice computation was done with $n_f=2$ degenerate light quarks with masses corresponding to an unphysical pion mass $\approx 640$~MeV and a lattice spacing $a^{-1}\approx 2.37$~GeV. Using the data for the ground and first excited states as well as the mixing angle in Eqs.~\eqref{s:match:e1}-\eqref{s:match:e2} one can obtain the lattice determination of the quarkonium static potential and the quarkonium-heavy-meson pair mixing potential. Similarly in Ref.~\cite{Bulava:2019iut} the static energies were studied in $n_f=2+1$ light-quarks and in addition to the states of Ref.~\cite{Bali:2005fu} the first strange heavy meson-antimeson-pair was included. Unfortunately, in Ref.~\cite{Bulava:2019iut} the mixing angles are not available and the static potentials cannot be extracted without heavy modeling. We show the original data of Ref.~\cite{Bali:2005fu} in Fig.~\ref{lat_data} and data transformed with Eqs.~\eqref{s:match:e1}-\eqref{s:match:e2} in Fig.~\ref{QQbpotplot}. It is interesting to note that the small bump in the first excited state (yellow triangles in Fig.~\ref{lat_data}), which in the range of $r$ where the bump occurs is dominated by the heavy meson-antimeson component, disappears in the transformed data for the heavy meson-antimeson static potential (yellow triangles in Fig.~\ref{QQbpotplot}). This seems to indicate that the short-distance heavy meson-antimeson interaction is a result of the mixing with quarkonium. Therefore, since the transformed data for the heavy meson-antimeson static potential is completely flat, our choice for the heavy meson-antimeson static potential in Eq.~\eqref{s:ths:e6} is consistent with the lattice data. Furthermore, the plot of the data for the mixing angle in the right-hand side of Fig.~\ref{lat_data} shows that the mixing angle is close to $0$ or $\pi/2$ except for a narrow region between $r\sim 1.2-1.3$~fm around the string-breaking distance. Therefore, the mixing potential can be considered a perturbation for most of the range of $r$.

In the following we focus on finding a parametrization of the quarkonium static potential. The quarkonium potential to be used in the Schrödinger equation is constructed combining the short-distance perturbative expression with the long-distance lattice data. For distances $r\leq 1$~GeV$^{-1}$ the static potential is set to the perturbative expression in the RS' scheme. The convergence of the $r$-dependence of the perturbative potential is poor at short distances if the renormalization scale is fixed~\cite{Kiyo:2010jm} due to the presence of large logarithms. If we set $\nu\sim 1/r$ the large logarithms associated with the soft scale are resummed into the running of $\alpha_s(\nu)$ and the convergence is improved. Specifically, we set $\nu=2/r$.

For distances $r> 1$~GeV$^{-1}$ the static potential is set to a fit of the lattice data. We parametrize the lattice data with the following function:
\begin{align}
V^{(0)}_{\rm L}(r)=\frac{b_1}{r}+\frac{b_2 r}{b_3 r+1}+b_4+\sigma r\,,\label{lat_pot}
\end{align}
where the linear coefficient is fixed to $\sigma=0.21$~GeV$^2$ to reproduce the long-range behavior found in Ref.~\cite{Luscher:2002qv}. We constrain the parameters so that the slope at the matching point $r_m=1$~GeV$^{-1}$ is equal to that of the perturbative potential. The rest of parameters are obtained by fitting the lattice data, we find
\begin{align}
b_1&=0.619\,,\\
b_2&=-1.774~{\rm GeV}^2\,,\\
b_3&=1.546~{\rm GeV}\,,\\
b_4&=-0.183~{\rm GeV}\,.
\end{align}
Due to possible powerlike divergences in the lattice computations the normalization of the lattice data is unknown. This ambiguity is removed by shifting $V^{(0)}_{\rm L}$ by a constant chosen to ensure that at the matching point the shifted lattice parametrization is continuous with the perturbative potential. Finally, the static potential we use in the Schrödinger equation reads as
\begin{align}
V_\Psi^{(0)}(r)=V_{\rm RS'}^{(0)}(\nu_f=0.7,\nu=2/r,r)\theta(r_m-r)+(V^{(0)}_{\rm L}(r)+\delta E_{\rm offset})\theta(r-r_m)\,,\label{full_pot}
\end{align}
with $\delta E_{\rm offset}=0.741$~GeV. We plot Eq.~\eqref{full_pot} in Fig.~\ref{QQbpotplot}.

\subsection{Mixing potential}\label{s:mixpot}

The lattice determination of the mixing potential is obtained using Eq.~\eqref{s:match:e3} and the data of Ref.~\cite{Bali:2005fu}. Since the mixing potential is proportional to the difference between the ground and first excited states it is not affected by the ambiguity in the normalization of the energies of the lattice computation. 

\begin{figure}[ht!]
\centerline{\includegraphics[width=0.7\linewidth]{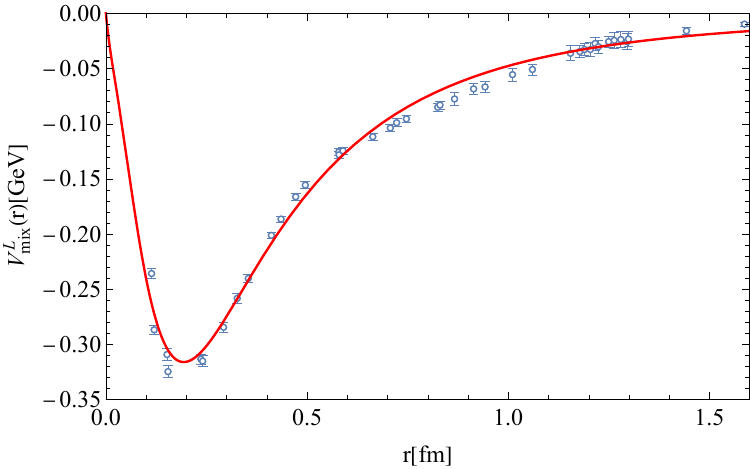}}
\caption{Quarkonium-heavy meson pair mixing potential. The blue open circles correspond to the lattice data of Ref.~\cite{Bali:2005fu} transformed into the mixing potential using Eq.~\eqref{s:match:e3}. The red line corresponds to the parametrization in Eq.~\eqref{mixpl} fitted to the lattice data.}
\label{mixpotplot}
\end{figure}

To parametrize the mixing potential we use a function that interpolates between the short- and long-distance behaviors as introduced in Ref.~\cite{Soto:2021cgk}. For distances $r\ll 1/\Lambda_{\rm QCD}$ the relative momentum between the heavy quarks can be integrated out and the quarkonium-heavy-meson pair mixing can be studied in weakly coupled pNRQCD~\cite{Pineda:1997bj,Brambilla:1999xf}. In this regime the heavy-quark fields can be decomposed into color-singlet and color-octet fields. At leading order in the multipole expansion the quarkonium states overlap with the singlet field while the heavy meson pair states can in principle overlap with both. The transition between the quarkonium state and the octet piece of the heavy meson pair is generated at leading order by a chromoelectric dipolar operator. The $\bm{r}$ factor in this operator provides the correct dependence on $\hat{\bm{r}}$ of the mixing operator for $\kappa^{pc}=1^{--}$ in Eq.~\eqref{sec:ths:e2} and produces a linear dependence of the mixing potential at leading order. Furthermore, the chromoelectric operator has the right quantum numbers to create the light-quark content of the heavy-meson pair. The second contribution corresponds to the overlap of the quarkonium state with the singlet piece of the heavy-meson pair. In this case the leading order transition is generated by three dipolar operators, therefore this contribution to the mixing potential has a $r^3$ dependence at leading order. This second contribution is in principle suppressed respect to the first one in the multipole expansion, however the size of the overlaps of the heavy meson pair state with the singlet and octet fields are unknown. For this reason we keep both contributions in our short-distance description of the mixing potential
\begin{align}
V^{\,{\rm (s.d.)}}_{\rm mix}(r)=c_1 r+c_2 r^3\,.  
\end{align}
For distances $r\gg 1/\Lambda_{\rm QCD}$ the mixing potential can be expanded in powers of $1/(\Lambda_{\rm QCD} r)^n$. If we assume that the mixing potential vanishes in the $r\to \infty$ limit then only $n\geq 0$ is allowed. By fitting the data with $r>1$~fm we find that the long-distance part is well described by
\begin{align}
V^{\,{\rm (l.d.)}}_{\rm mix}(r)=\frac{c_3}{r^3}\,.  
\end{align}
We construct the interpolation by summing the short- and long-distance descriptions multiplied by interpolating functions depending on $r$ and a new $r_0$ parameter. The interpolating functions are $w_s=(r_0/(r+r_0))^n$ and $w_l=(r/(r+r_0))^n$ for the short- and long-distance pieces, respectively. The $r_0$ parameter determines the value of $r$ where both interpolating functions are equal. We pick $r_0=0.25$~fm as it is a reasonable point for the breakdown of the multipole expansion. The full parametrization of the potential is as follows
\begin{align}
V^{\rm L}_{\rm mix}(r)=w_s(r) V^{\,{\rm (s.d.)}}_{\rm mix}(r)+w_l(r)V^{\,{\rm (l.d.)}}_{\rm mix}(r)\,.\label{mixpl}
\end{align}
The value of $n$ is set to the smallest value that the short- and long-distance potentials dominate in their respective limits, which in this case is $n=7$. The rest of the parameters are fitted to the lattice data, we find
\begin{align}
c_1&=-0.723~{\rm GeV}^2\,,\\
c_2&=-15.251~{\rm GeV}^4\,,\\
c_3&=-13.991~{\rm GeV}^{-2}\,.
\end{align}
It is interesting that the value of $c_2$ is not as suppressed with respect to the one of $c_1$ as one would expect from the pNRQCD counting which might indicate that the heavy meson pair has a larger overlap with the singlet field than the octet one in the short-distance regime. This is consistent with the slightly attractive behavior of the first excited static state (in yellow triangles in Fig.~\ref{lat_data}) in the first few short-distance data points\footnote{In Ref.~\cite{Bali:2005fu} the attractive nature of the short-distance data of the first excited static state was already linked with a dominating overlap with a heavy quark-antiquark singlet state.}. In Fig.~\ref{mixpotplot} we plot the lattice data for the mixing potential and the parametrization in Eq.~\eqref{mixpl}. 

The normalization of the mixing operator in Ref.~\cite{Bali:2005fu} and ours in the Lagrangian in Eq.~\eqref{sec:ths:e2} for $\kappa=1$ coincide, however the heavy meson-antimeson pair interpolating operator is isospin $I=0$ unlike ours, in Eq.~\eqref{sec:mnrq:e2}, which corresponds to a single light-quark flavor. In other words, the heavy meson-antimeson pair operator in Ref.~\cite{Bali:2005fu} interpolates for a field in BOEFT corresponding to the normalized sum of the charged and neutral heavy meson-antimeson pairs. Since the two lattice light quarks are degenerate we have
\begin{align}
V^{1}_{\rm mix}(r)=\frac{1}{\sqrt{2}}V^{\rm L}_{\rm mix}(r)\,.\label{mpp}
\end{align}
We will use this mixing potential for all three light-quark flavors, which is an approximation. However, the light-quark mass used in Ref.~\cite{Bali:2005fu} is in fact closer to the strange mass than the up or down masses, therefore it can be expected that the approximation produces more accurate results for pairs of heavy mesons with strangeness.

In Ref.~\cite{Bulava:2019iut} the static energy spectrum of quarkonium coupled to heavy meson pairs without and with strangeness was obtained in lattice QCD for distances between $r\sim 0.25~{\rm fm}$ and  $r\sim 1.6~{\rm fm}$. Since the mixing angles were not given the mixing potentials can only be extracted by assuming specific forms of the quarkonium and heavy meson pair static potentials. Assuming a Cornell type potential for the former and a constant for the latter one can fit the mixing potential. Since only the long-distance regime is available we fitted mixing potentials $\sim r^{-1}$, $\sim r^{-3}$ and a constant, the latter being the choice in Ref.~\cite{Bulava:2019iut} analysis. In all three cases the fits were of similar quality. Therefore, we conclude that a reliable estimation of the mixing potentials from the data of Ref.~\cite{Bulava:2019iut} is not possible.

The mixing potential has been extracted from the lattice data of Ref.~\cite{Bali:2005fu} in Refs.~\cite{Bicudo:2019ymo,Bicudo:2020qhp,Bicudo:2022ihz}. However a different procedure was followed. It was argued that the heavy-meson pair to heavy-meson pair correlator of Ref.~\cite{Bali:2005fu}, interpolates not only for the $\Sigma_g^+$ representation but also for $\Pi_g$ and $\Sigma_u^-$ representations. As a result the authors of Refs.~\cite{Bicudo:2019ymo,Bicudo:2020qhp,Bicudo:2022ihz} argue that lattice data should be fitted with a parametrization that takes into account these extra states in the meson pair to meson pair correlator. Lets us note, that fits to the correlators with extra states where also considered in Ref.~\cite{Bali:2005fu} but where found not to describe the data well. Since the original data on the correlators of Ref.~\cite{Bali:2005fu} is not available, the authors of Refs.~\cite{Bicudo:2019ymo,Bicudo:2020qhp,Bicudo:2022ihz} resample the lattice correlators using the original parametrization and then fit this resampled data with their parametrization containing the extra states. While we agree on the initial point about the extra states in the meson pair to meson pair correlator, we do not think the resampled data can contain information on these extra states as it was produced from the original parametrization. The quarkonium static potential obtained from the resampled data can be found in Fig.~3 of Ref.~\cite{Bicudo:2019ymo}. If we compare it to the one we obtain, in Fig.~\ref{QQbpotplot}, it can be observed that the shapes are notably different. In our determination, the shape of the static potential is closer to previous lattice determinations of the static potential that did not include the mixing with the threshold, which is the behavior expected away from the string breaking distance as we discussed at the end of Sec.~\ref{sec:mnrq}. A possible explanation for the extra states in the meson pair to meson pair correlator of Ref.~\cite{Bali:2005fu} not showing up in their fits is that the extra states are degenerate with the $\Sigma_g^+$ one, as we do in the Lagrangian in Eq.~\eqref{s:ths:e6}. Therefore, in our opinion the most appropriate way of extracting the static and mixing potentials is to use the two state parametrizations of the lattice correlators. We hope that in the future new lattice studies clarify this issue.

In Refs.~\cite{Bruschini:2020voj,Bruschini:2021sjh} a model for the mixing potential is used. This consists of a Gaussian shape with the maximum at the string breaking distance, i.e. the value of $r$ when $V^{(0)}_{\Psi}(r_{\rm s.b.})=V^{(0)}_{1^{--}0}(r_{\rm s.b.})$. Comparing with the mixing potential extracted from the lattice data in Fig.~\ref{mixpotplot} we can see that the model misses important features. The maximum of the mixing potential occurs at $r\sim 0.25$~fm instead of the string breaking distance and as a result the overall shape is different. Moreover, as the value of the mixing potential
at the string breaking distance must be equal to half the avoided crossing separation\footnote{The avoided crossing distance is $(V^{(0)}_{2\Sigma_g^+}(r_{\rm s.b.})-V^{(0)}_{1\Sigma_g^+}(r_{\rm s.b.}))$.},
the maximum value of the model mixing potential of Refs.~\cite{Bruschini:2020voj,Bruschini:2021sjh} is much smaller than the one of the potential extracted from the lattice data.

\subsection{ \texorpdfstring{$1/m_Q$}{1/mQ} quarkonium potential}

To improve the accuracy in the determination of the quarkonium spectrum we compute the contribution of the $1/m_Q$ suppressed potential~\cite{Brambilla:2000gk} using standard time independent perturbation theory. To obtain an expression for this potential we follow an analogous approach to the one in Sec.~\ref{qssp} for the static potential. We construct the potential by combining the short-distance perturbative expression with the available lattice data for long-distances. We use the leading order perturbative result from Ref.~\cite{Peset:2017wef}. It reads as
\begin{align}
V_{\rm p.t.}^{(1)}(r)=-\frac{\alpha^2_s(\nu)C_AC_F}{4r^2}\,.
\label{1omptpot}
\end{align}
Notice, that the form of the potential depends on the matching scheme. We use the expression for the Wilson loop matching scheme in accordance with the rest of the paper. As in the static potential, we resumme large soft logs by setting $\nu=2/r$.

\begin{figure}[ht!]
\centerline{\includegraphics[width=0.7\linewidth]{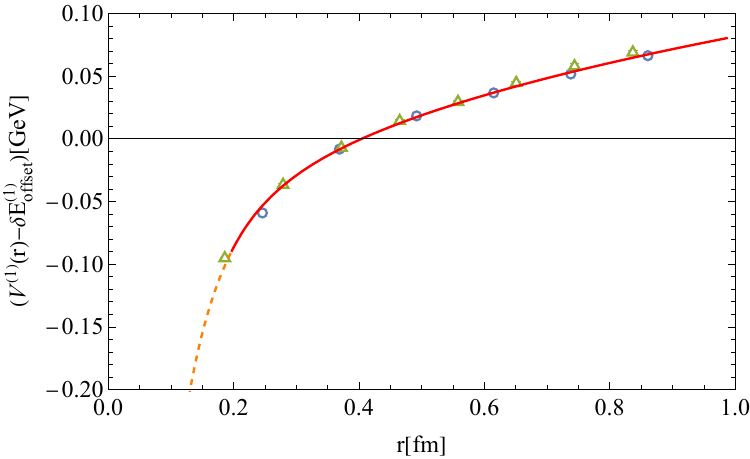}}
\caption{Plot of the $1/m_Q$-suppressed quarkonium potential given in Eq.~\eqref{full_1om_pot} with parameters fitted as described in the text. The orange dashed line corresponds to the perturbative part in Eq.~\eqref{1omptpot} while the red continuous line corresponds to the parametrization of the lattice data in Eq.~\eqref{1omlatpot}. The blue circles and green triangles corresponds to the lattice data with lattice coupling $\beta=5.85$ and $\beta=6.00$, respectively, from Refs.~\cite{Koma:2006si,Koma:2007jq}.}
\label{1ompotplot}
\end{figure}

The $1/m_Q$ suppressed quarkonium potential has been studied in the lattice in Refs.~\cite{Koma:2006si,Koma:2007jq,Koma:2012bc}. We use two datasets with simulation parameters $\beta=5.85$, $a=0.123$~fm and $\beta=6.00$, $a=0.093$~fm in the quenched approximation. The lattice data is plotted in Fig.~\ref{1ompotplot}. To parametrize the lattice data we use the following function
\begin{align}
V^{(1)}_{\rm L}=-\frac{d_1}{r^2}+\frac{d_2 r}{d_3r+1}+d_4+\sigma_1 \log r\,,\label{1omlatpot}
\end{align}
which interpolates between the dependence on $r^{-2}$ from the perturbative expression in Eq.~\eqref{1omptpot} and the $\log r$ dependence obtained from Effective String Theory~\cite{Nambu:1978bd,Luscher:1980fr,Luscher:2002qv,Perez-Nadal:2008wtr,Brambilla:2014eaa,Hwang:2018rju,Soto:2021cgk}\footnote{Effective String Theory is successful dynamical model for the long-distance regime (i.e., $r\gg 1/\Lambda_{\rm QCD}$) and provides expressions for the quarkonium potentials in powers of $\Lambda_{\rm QCD}/r$ in accordance to our argument for the mixing potential in Sec.~\ref{s:mixpot}.}. The parameters of Eq.~\eqref{1omlatpot} are constrained to reproduce the slope of the perturbative potential at the matching point $r_m=1$~GeV$^{-1}$, the rest of the parameters are fitted to the lattice data. The values we obtained are as follows:
\begin{align}
d_1&=0.114\,,\\
d_2&=-7.704~{\rm GeV}^3\,,\\
d_3&=5.823~{\rm GeV}\,,\\
d_4&=1.149~{\rm GeV}^2\,,\\
\sigma_1&=0.129~{\rm GeV}^2\,.
\end{align}
The ambiguity in the normalization of the lattice data is removed by shifting $V^{(1)}_{\rm L}$ so that the value at the matching point is equal to that of the perturbative expression in Eq.~\eqref{1omptpot}. The full expression of the $1/m_Q$ suppressed potential is 
\begin{align}
V_\Psi^{(1)}(r)=V_{\rm p.t.}^{(1)}(r)\theta(r_m-r)+(V^{(1)}_{\rm L}(r)+\delta E^{(1)}_{\rm offset})\theta(r-r_m)\,,\label{full_1om_pot}
\end{align}
with $\delta E^{(1)}_{\rm offset}=-0.088$~GeV. In Fig.~\ref{1ompotplot} we plot the potential in Eq.~\eqref{full_1om_pot}. We should note that the value of $\delta E^{(1)}_{\rm offset}$ depends noticeably on the specific matching  point $r_m$ and the order at which we take the perturbative potential in Eq.~\eqref{1omptpot}. This is a result of the lack of lattice data at shorter distances and the possible existence of renormalon ambiguities in $V_{\rm p.t.}^{(1)}(r)$.

\subsection{Numerical results}\label{sec:nr}

We solve numerically the Schrödinger equation for the quarkonium static potential in Eq.~\eqref{full_pot}
\begin{align}
\left[-\frac{\bnabla^2}{m_Q}+V_\Psi^{(0)}(r)\right]\Psi_{nl}(\bm{r})=E_{nl}^{(0)}\Psi_{nl}(\bm{r})\label{sec:nr:e1}\,,
\end{align}
obtaining the wave functions $\Psi_{nl}$ and eigenenergies $E_{nl}^{(0)}$, with $n$ and $l$ the principal and angular quantum numbers of the quarkonium state, respectively. Using the wave functions we compute the contribution of the $1/m_Q$ suppressed potential in Eq.~\eqref{full_1om_pot} in standard quantum mechanical perturbation theory.
\begin{align}
E_{nl}^{(1)}=\int d^3\bm{r}\, \Psi^*_{nl}(\bm{r})V_{\Psi}^{(1)}(r)\Psi_{nl}(\bm{r})
\end{align}
The results for $E_{nl}^{(0)}$ and $E_{nl}^{(1)}$ can be found in Tables~\ref{sp_bb_s}-\ref{sp_bb_d} for bottomonium states and Tables~\ref{sp_cc_s}-\ref{sp_cc_d} for charmonium states. The uncertainties in these two quantities are estimated by their difference when changing the matching point between the perturbative expressions and the lattice data fits from $r_m=1$~GeV$^{-1}$ to $r_m=0.66$~GeV$^{-1}$.

At the order we are working, isospin and heavy-quark spin contributions can be neglected, thus we use heavy meson masses reflecting these approximations. The values of the heavy meson masses are obtained by first computing the average of the neutral and charge states, when both are available, and then the spin average of the scalar and vector states. For strange heavy mesons only the last step is necessary. The masses of the physical heavy meson states are taken from the PDG~\cite{ParticleDataGroup:2020ssz}. We find
\begin{align}
m_D&=1.97322~{\rm GeV}\,,\\
m_B&=5.3134~{\rm GeV}\,,\\
m_{D_s}&=2.0762~{\rm GeV}\,,\\
m_{B_s}&=5.4033~{\rm GeV}\,.
\end{align}

The heavy meson-antimeson pair contributions to the quarkonium masses are computed using Eq.~\eqref{mw:e4}. For this computation we take the quarkonium binding energy up to next-to-leading order. Since the results in this section are all for heavy meson-antimeson pairs with $\kappa^{pc}=1^{--}$ we will drop this label. On the other hand, since we will compute the contributions for heavy mesons without and with strangeness we  will add a label $f$ indicating the light-quark flavor. We use $f=q$ to denote the sum of the neutral and charged heavy meson-antimeson pair contributions. These two contributions are equal in the isospin limit. To denote the contributions of heavy mesons pairs with closed strangeness we use $f=s$. The vertex form factor $a^{l'}_{nl}(k)$ is computed numerically from the expression in Eq.~\eqref{mw:e6} using the wave functions form solving Eq.~\eqref{sec:nr:e1} and the mixing potential in Eq.~\eqref{mpp}. We sample it for a range of $k$ from $0$~GeV to $6$~GeV at $10$~MeV intervals. A linear interpolation of this data is then used to compute the mass contribution in Eq.~\eqref{mw:e4}. The uncertainty of $E^{fl'}_{nl}$ is estimated as follows. The uncertainty of $k^2_d$ is obtained by combining in quadrature the uncertainties of $E_{nl}^{(0)}$, $E_{nl}^{(1)}$, $m^{\rm RS'}_Q$ and the size of higher order contributions ${\cal O}(\Lambda^2_{\rm QCD}/m^2_c)\sim 40$~MeV, ${\cal O}(\Lambda^2_{\rm QCD}/m^2_b)\sim 4$~MeV. The mass contribution is computed for a random Gaussian sample of $k^2_d$ with the standard deviation set to its uncertainty. The average and standard deviation of the set of results for $E^{fl'}_{nl}$ are taken as our central value and its uncertainty, respectively. The contributions of the heavy meson pairs to the quarkonium spectrum is displayed in Tables~\ref{sp_bb_s}-\ref{sp_bb_d} for bottomonium and Tables~\ref{sp_cc_s}-\ref{sp_cc_d} for charmonium.

The total masses of the quarkonium states are obtained as
\begin{align}
M_{nl}=2m^{\rm (RS')}_Q+E_{nl}^{(0)}+E_{nl}^{(1)}+\sum_{f=q,s}\sum^{l+1}_{l'=|l-1|}E^{fl'}_{nl}\,.\label{s:nr:e2}
\end{align}
The uncertainty of $M_{nl}$ is obtained by adding in quadrature the uncertainties of each term in Eq.~\eqref{s:nr:e2}.

\begin{table}[ht!]
\begin{tabular}{cccccc} \hline\hline
$n$ & $E_{nS}^{(0)}$ & $E_{nS}^{(1)}$ & $E^{qP}_{nS}$  & $E^{sP}_{nS}$ & $M_{nS}$ \\ \hline
$1$	& $-291(7)$	 & $-54(1)$  & $-42(2)$	& $-19(1)$ & $9491(90)$ \\
$2$	& $168(7)$	 & $-28(2)$  & $-27(2)$	& $-11(1)$ & $9999(90)$ \\
$3$	& $492(8)$	 & $-20(2)$  & $-21(2)$	& $-8(1)$	 & $10341(90)$ \\
$4$	& $767(9)$	 & $-15(1)$  & $-19(6)$	& $-7(1)$	 & $10623(90)$ \\
$5$	& $1015(11)$ & $-11(1)$  & $-15(3)$	& $-6(1)$	 & $10881(90)$ \\
$6$	& $1244(12)$ & $-8(1)$	 & $-15(5)$	& $-5(1)$	 & $11114(90)$ \\ \hline\hline
\end{tabular}
\caption{Spectrum of $S$-wave bottomonium states. All entries in MeV.}
\label{sp_bb_s}
\end{table}

\begin{table}[ht!]
\begin{tabular}{cccccccc} \hline\hline
$n$ & $E_{nP}^{(0)}$ & $E_{nP}^{(1)}$ & $E_{nP}^{qS}$ & $E_{nP}^{sS}$ & $E_{nP}^{qD}$ & $E_{nP}^{qD}$ & $M_{nP}$ \\ \hline
$1$	& $59(10)$	 & $-21(2)$ & $-22(2)$	& $-9(1)$	& $-27(2)$ & $-12(1)$	 & $9867(90)$ \\
$2$	& $389(8)$	 & $-14(2)$ & $-15(2)$	& $-6(1)$	& $-16(2)$ & $-6.8(5)$ & $10229(90)$ \\
$3$	& $671(9)$	 & $-10(2)$ & $-13(4)$	& $-5(1)$	& $-14(3)$ & $-5.1(4)$ & $10522(90)$ \\
$4$	& $924(10)$	 & $-7(1)$	& $-10(4)$	& $-5(1)$	& $-9(2)$	 & $-5(1)$	 & $10786(90)$ \\
$5$	& $1156(11)$ & $-4(2)$	& $-10(3)$	& $-3(1)$	& $-8(2)$	 & $-3.1(5)$ & $11026(90)$ \\
$6$	& $1375(13)$ & $-2(1)$	& $-10(2)$	& $-3(1)$	& $-8(2)$	 & $-2.9(4)$ & $11246(90)$ \\ \hline\hline
\end{tabular}
\caption{Spectrum of $P$-wave bottomonium states. All entries in MeV.}
\label{sp_bb_p}
\end{table}

\begin{table}[ht!]
\begin{tabular}{cccccc} \hline\hline
$n$ & $E_{nD}^{(0)}$ & $E_{nD}^{(1)}$ & $E_{nD}^{qP}$ & $E_{nD}^{sP}$ & $M_{nD}$ \\ \hline
$1$	& $274(8)$	 & $-14(2)$ & $-18(2)$	& $-7(1)$	& $10133(90)$ \\
$2$	& $565(9)$	 & $-10(1)$ & $-17(3)$	& $-6(1)$	& $1043(90)$ \\
$3$	& $824(10)$	 & $-6(2)$	& $-15(5)$	& $-5(1)$	& $10696(90)$ \\
$4$	& $1062(12)$ & $-3(2)$	& $-15(4)$	& $-5(1)$	& $10937(90)$ \\
$5$	& $1284(13)$ & $-1(2)$	& $-15(3)$	& $-5(1)$	& $11161(90)$ \\
$6$	& $1494(14)$ & $0(0)$	  & $-14(2)$	& $-5(1)$	& $11374(91)$ \\ \hline\hline
\end{tabular}
\caption{Spectrum of $D$-wave bottomonium states. All entries in MeV.}
\label{sp_bb_d}
\end{table}

\begin{table}[ht!]
\begin{tabular}{cccccc} \hline\hline
$n$ & $E_{nS}^{(0)}$ & $E_{nS}^{(1)}$ & $E_{nS}^{qP}$ & $E_{nS}^{sP}$ & $M_{nS}$ \\ \hline
$1$	& $44(9)$	   & $-79(5)$ & $-31(2)$  & $-14(1)$  & $3104(90)$ \\
$2$	& $606(8)$	 & $-42(4)$ & $-17(3)$  & $-7(1)$	  & $3726(90)$ \\
$3$	& $1051(11)$ & $-22(5)$ & $-12(3)$  & $-5(1)$	  & $4195(90)$ \\
$4$	& $1440(13)$ & $-8(5)$	& $-4(3)$	  & $-5(1)$	  & $4607(91)$ \\
$5$	& $1793(15)$ & $3(6)$	  & $-1(1)$	  & $-2(1)$	  & $4979(91)$ \\
$6$	& $2122(17)$ & $12(6)$	& $-0.1(4)$ & $-0.4(6)$ & $5318(91)$ \\
\hline\hline
\end{tabular}
\caption{Spectrum of $S$-wave charmonium states. All entries in MeV.}
\label{sp_cc_s}
\end{table}

\begin{table}[ht!]
\begin{tabular}{cccccccc} \hline\hline
$n$ & $E_{nP}^{(0)}$ & $E_{nP}^{(1)}$ & $E_{nP}^{qS}$ & $E_{nP}^{sS}$ & $E_{nP}^{qD}$ & $E_{nP}^{sD}$ & $M_{nP}$ \\ \hline
$1$	& $415(8)$	 & $-36(6)$ & $-15(3)$ & $-5(1)$  & $-13(1)$  & $-5.5(4)$ & $3525(90)$ \\
$2$	& $880(11)$	 & $-17(5)$ & $-24(5)$ & $-7(2)$  & $-10(2)$	& $-4(1)$	  & $4003(90)$ \\
$3$	& $1282(13)$ & $-3(6)$  & $-4(5)$	 & $-7(1)$  & $-5.3(4)$ & $-3.1(3)$ & $4444(91)$ \\
$4$	& $1645(14)$ & $7(6)$	  & $3(2)$   & $-1(1)$  & $-4(1)$	  & $-2.1(1)$ & $4832(91)$ \\
$5$	& $1982(17)$ & $16(6)$	& $4(1)$	 & $1(1)$	  & $-4(1)$	  & $-1.8(2)$ & $5181(91)$ \\
$6$	& $2298(19)$ & $23(6)$	& $4.3(4)$ & $1.6(5)$ & $-4(1)$	  & $-1.6(2)$ & $5505(92)$ \\
\hline\hline
\end{tabular}
\caption{Spectrum of $P$-wave charmonium states. All entries in MeV.}
\label{sp_cc_p}
\end{table}

\begin{table}[ht!]
\begin{tabular}{cccccc} \hline\hline
$n$ & $E_{nD}^{(0)}$ & $E_{nD}^{(1)}$ & $E_{nD}^{qP}$ & $E_{nD}^{sP}$ & $M_{nD}$ \\ \hline
$1$	& $696(8)$	 & $-20(5)$ & $-16(7)$	 & $-4(1)$	   & $3840(90)$  \\
$2$	& $1115(11)$ & $-5(6)$	& $-10(6)$	 & $-7(2)$	   & $4276(91)$ \\
$3$	& $1490(14)$	 & $6(6)$	  & $0.1(3.4)$ & $-4(2)$	   & $4675(91)$ \\
$4$	& $1835(16)$ & $15(7)$	& $3(2)$	   & $-0.4(1.5)$ & $5037(91)$ \\
$5$	& $2158(18)$ & $22(6)$	& $4(1)$	   & $1(1)$	     & $5369(91)$ \\
$6$	& $2463(19)$ & $29(7)$	& $4.2(4)$	 & $1.7(4)$	   & $5682(92)$ \\
\hline\hline
\end{tabular}
\caption{Spectrum of $D$-wave charmonium states. All entries in MeV.}
\label{sp_cc_d}
\end{table}

Our result show the contribution of these two thresholds to the quarkonium masses is comparable to that of the $1/m_Q$ suppressed potential. Contrary to what it could be expected intuitively, the contribution of the thresholds is larger for the lower lying quarkonium states that the excited ones. The underlying reason is in the shape on the mixing potential (see Fig.~\ref{mixpotplot}) which peaks, in absolute value, at $r\sim 0.25$~fm, while the excited states wave functions extend to far longer ranges. 

\begin{figure}[ht!]
\centerline{\includegraphics[width=0.8\linewidth]{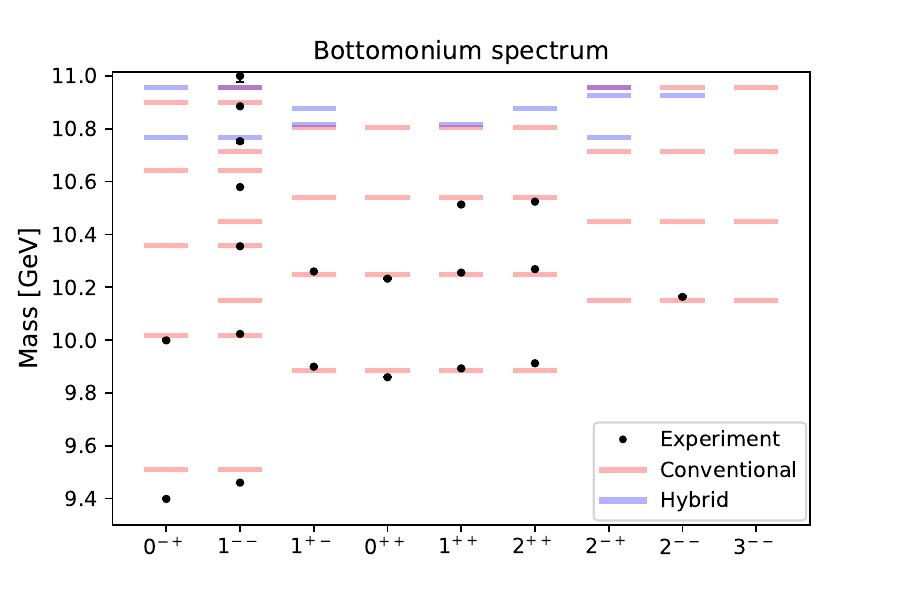}}
\caption{Comparison of the experimental bottomonium spectrum (black dots) with the spectrum we have obtained (red lines). We also include the hybrid bottomonium states (blue lines) from Ref.~\cite{Berwein:2015vca} in the mass range and $J^{PC}$ of the figure. Both conventional and hybrid bottomonium spectra are shifted so the $2S$ state mass matches the experimental spin average one.}
\label{plt_bbbar_spc}
\end{figure}

\begin{figure}[ht!]
\centerline{\includegraphics[width=0.8\linewidth]{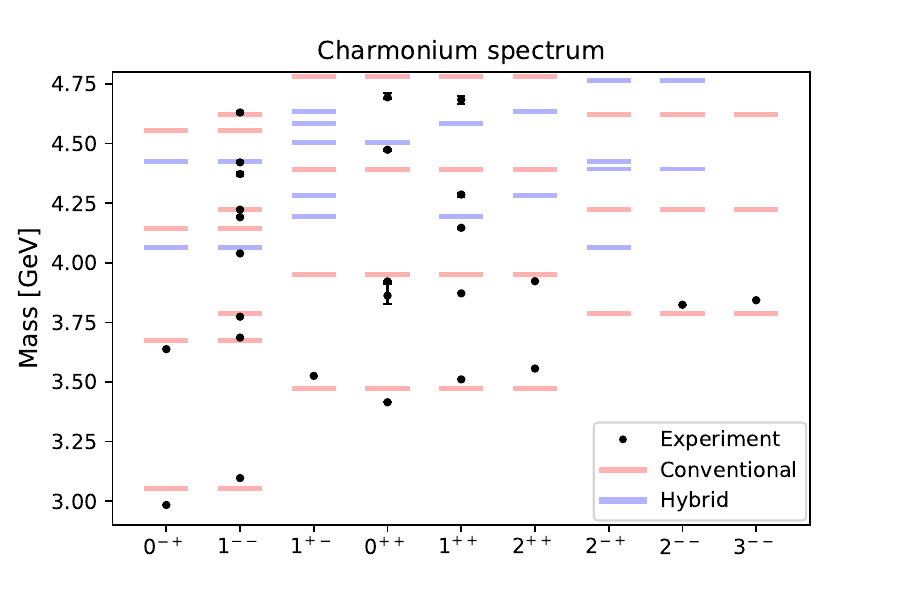}}
\caption{Comparison of the experimental charmonium spectrum (black dots) with the spectrum we have obtained (red lines). We also include the hybrid charmonium states (blue lines) from Ref.~\cite{Berwein:2015vca} in the mass range and $J^{PC}$ of the figure. Both conventional and hybrid charmonium spectra are shifted so the $2S$ state mass matches the experimental spin average one.}
\label{plt_ccbar_spc}
\end{figure}

The uncertainty of our results for the quarkonium masses in Tables~\ref{sp_bb_s}-\ref{sp_cc_d} is dominated by the uncertainty in the determination of the heavy quark masses in the RS' scheme. This source of uncertainty cancels out in mass differences which are consequently much more accurate. Furthermore, we can reconstruct the spectrum by adding to the experimental mass of a given state our mass differences with respect to the same state. Since our computation does not include spin-dependent contributions it is convenient to consider as a reference an $S$-wave state, since the spin-averages of these are independent of such contributions. Additionally, we expect the spin-independent ${\cal O}(1/m^2_Q)$ contributions to be smaller for higher excited states, in a similar way as the spin-independent ${\cal O}(1/m_Q)$ contribution we have computed. Therefore, we choose as experimental reference state the $2S$ doublet, since this is the higher laying $S$-wave doublet which has been measured for both charmonium and bottomonium. In Table~\ref{sp_sft} we show the bottomonium and charmonium spectra shifted so the $2S$ state mass matches the spin-average of the corresponding experimental masses. Both shifts are within the uncertainty of the heavy quark masses. In Figs.~\ref{plt_bbbar_spc} and \ref{plt_ccbar_spc} we show all the experimental bottomonium and charmonium states listed in the PDG with definite $J^{PC}$~\cite{ParticleDataGroup:2020ssz} compared to the shifted spectra. We also display the hybrid quarkonium states expected to appear in the energy range of the figures and with $J^{PC}$ matching those allowed for $S,P$ and $D$ wave quarkonium. However, one should keep in mind that exotic $J^{PC}$ are possible for quarkonium hybrids, including, for instance, heavy-quark spin partners of the states displayed. The mass values of the hybrid quarkonium are taken from Ref.~\cite{Berwein:2015vca} and also shifted to match the experimental value of spin-average mass of the $2S$ doublet. To do this, we obtain the $2S$ state mass for the $\Sigma_g^+$ static energy from the lattice data of Ref.~\cite{Juge:2002br}, which is the same source as for the $\Pi_u-\Sigma_u^-$ static energy data, used in Ref.~\cite{Berwein:2015vca} to obtain the hybrid spectra. It should be kept in mind that the hybrid quarkonium states displayed in Figs.~\ref{plt_bbbar_spc} and \ref{plt_ccbar_spc} are not computed to the same accuracy as the conventional quarkonium ones, since they do not include $O(1/m_Q)$ corrections\footnote{Spin-dependent contributions appear at order $O(1/m_Q)$ for hybrid quarkonium~\cite{Brambilla:2019jfi}.} nor heavy meson-antimeson pair contributions.

\begin{table}[ht!]
\begin{minipage}[t]{0.4\linewidth}
\begin{tabular}{c|ccc} \hline\hline
\backslashbox{n}{l} & $0$     & $1$   & $2$ \\ \hline
$1$	& $9.509(8)$	& $9.885(11)$	& $10.151(9)$ \\
$2$	& $10.017({\rm e})$	& $10.248(10)$	& $10.448(10)$ \\
$3$	& $10.359(9)$	& $10.540(11)$	& $10.714(12)$ \\
$4$	& $10.641(12)$	& $10.804(12)$	& $10.955(13)$ \\
$5$	& $10.899(12)$	& $11.044(12)$	& $11.179(14)$ \\
$6$	& $11.132(14)$	& $11.264(14)$	& $11.392(15)$ \\
\hline\hline
\end{tabular}
\end{minipage}
\begin{minipage}[t]{0.4\linewidth}
\begin{tabular}{c|ccc} \hline\hline
\backslashbox{n}{l} & $0$     & $1$   & $2$ \\ \hline
$1$	& $3.052(38)$	& $3.473(38)$	& $3.788(39)$ \\
$2$	& $3.67395({\rm e})$	& $3.951(39)$	& $4.224(39)$ \\
$3$	& $4.143(39)$	& $4.392(40)$	& $4.623(40)$ \\
$4$	& $4.555(40)$	& $4.780(40)$	& $4.985(41)$ \\
$5$	& $4.927(40)$	& $5.129(41)$	& $5.317(42)$ \\
$6$	& $5.266(41)$	& $5.453(42)$	& $5.630(42)$ \\
\hline\hline
\end{tabular}
\end{minipage}
\caption{Bottomonium (left) and charmonium (right) spectra with the origin of energies adjusted to the experimental value of the spin average of the respective $2S$ states. The experimental spin-average mass is marked by (e) and taken from the PDG~\cite{ParticleDataGroup:2020ssz}.}
\label{sp_sft}
\end{table}

\begin{figure}
\centerline{\includegraphics[width=0.4\linewidth]{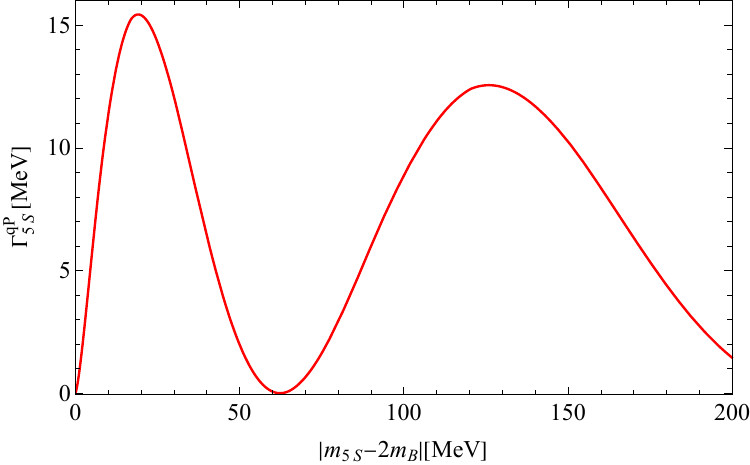}\includegraphics[width=0.4\linewidth]{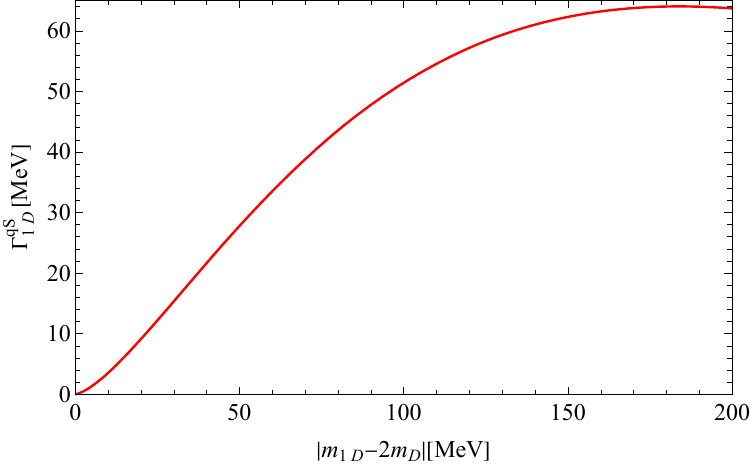}}
\caption{Plots of the decay width dependence with the mass difference between the quarkonium and the two meson masses. In the left we plot the width of the $5S$ bottomonium state decaying to $B\overline{B}$. In the right we plot the width of the $1D$ bottomonium state decaying to $D\overline{D}$. From the plots we can see that for variations of the mass difference of the order of its uncertainty the values of the widths can change by large amounts.}
\label{width_dep}
\end{figure}

The widths are computed according to Eq.~\eqref{mw:e5}. We take Eq.~\eqref{s:nr:e2} as the input mass. As in the mass computation, we compute the uncertainty $k^2_d$ adding up in quadrature the uncertainties of each term and the size of higher order contributions. We create a random Gaussian sample of $k^2_d$ values and compute the decay widths for each value in the set. The average and standard deviation of these computations are assigned as the central value and uncertainty of the widths. The notation for the widths follows the one for the energy contributions, in particular recall that $f=q$ corresponds to the sum of the widths for the neutral and charged heavy meson pairs. The results for the widths are shown in Tables~\ref{wdts_bb_S}-\ref{wdts_bb_D} for bottomonium states and Tables~\ref{wdts_cc_S}-\ref{wdts_cc_D} for charmonium states. We find values of about $5-10$~MeV for bottomonium and $10-50$~MeV for charmonium. The threshold contributions to the mass and width of $P$-wave quarkonia turn out to be slightly larger than $S$ and $D$ wave quarkonia due to it coupling to the heavy meson-antimeson pairs in two partial wave channels instead of one. As it can be seem from these results the uncertainties are large, particularly in comparison to the uncertainties in the mass contributions. This is a result on a strong dependence of the widths on the value of $k^2_d$. To illustrate this we plot the value of width as a function of the difference between the quarkonium and the heavy meson pair mass for two specific cases in Fig.~\ref{width_dep}. We can see that a large range of values of the widths can be produced within the uncertainty of the mass difference.

\begin{table}[ht!]
\begin{minipage}[t]{0.32\linewidth}
\begin{tabular}{cccc} \hline\hline
$n$ & $\Gamma^{\rm total}_{nS}$ & $\Gamma_{nS}^{qP}$ & $\Gamma_{nS}^{sP}$ \\ \hline
$5$	& $6(6)$	& $3(3)$	& $3(3)$ \\
$6$	& $4(7)$	& $2(6)$	& $1(1)$ \\ \hline\hline
\end{tabular}
\caption{Widths of $S$-wave bottomonium states.  All entries in MeV units.}
\label{wdts_bb_S}
\end{minipage}
\hfill
\begin{minipage}[t]{0.32\linewidth}
\begin{tabular}{cccccc} \hline\hline
$n$ & $\Gamma^{\rm total}_{nP}$ & $\Gamma_{nP}^{qS}$ & $\Gamma_{nP}^{sS}$ & $\Gamma_{nP}^{qD}$ & $\Gamma_{nP}^{sD}$ \\ \hline
$4$	& $6(7)$	& $3(4)$	& b.t.	 & $3(3)$	& b.t.    \\
$5$	& $6(5)$	& $2(2)$	& $2(1)$ & $1(1)$	& $1(1)$ \\
$6$	& $7(7)$	& $4(4)$	& $1(1)$ & $2(2)$	& $0.4(3)$ \\
\hline\hline
\end{tabular}
\caption{Widths of $P$-wave bottomonium states. All entries in MeV units. b.t. stands for below threshold.}
\label{wdts_bb_P}
\end{minipage}
\hfill
\begin{minipage}[t]{0.32\linewidth}
\begin{tabular}{cccc} \hline\hline
$n$ & $\Gamma^{\rm total}_{nD}$ & $\Gamma_{nD}^{qP}$ & $\Gamma_{nD}^{sP}$ \\ \hline
$3$	& $3(5)$   & $3(5)$	  & b.t.        \\
$4$	& $4(5)$	 & $2(4)$	  & $1(1)$     \\
$5$	& $7(8)$	 & $6(7)$	  & $1(1)$     \\
$6$	& $13(11)$ & $12(10)$	& $1(1)$ \\  \hline\hline
\end{tabular}
\caption{Widths of $D$-wave bottomonium states. All entries in MeV units. b.t. stands for below threshold.}
\label{wdts_bb_D}
\end{minipage}
\end{table}

\begin{table}[ht!]
\begin{minipage}[t]{0.32\linewidth}
\begin{tabular}{cccc} \hline\hline
$n$ & $\Gamma^{\rm total}_{nS}$ & $\Gamma_{nS}^{qP}$ & $\Gamma_{nS}^{sP}$ \\ \hline
$3$	& $16(10)$ & $15(8)$	& $1(2)$ \\
$4$	& $22(5)$	 & $16(2)$	& $6(3)$ \\
$5$	& $15(4)$	 & $9(3)$	  & $6(1)$ \\
$6$	& $9(3)$	 & $4(2)$	  & $4(1)$ \\  \hline\hline
\end{tabular}
\caption{Widths of $S$-wave charmonium states. All entries in MeV units.}
\label{wdts_cc_S}
\end{minipage}
\hfill
\begin{minipage}[t]{0.32\linewidth}
\begin{tabular}{cccccc} \hline\hline
$n$ & $\Gamma^{\rm total}_{nP}$ & $\Gamma_{nP}^{qS}$ & $\Gamma_{nP}^{sS}$ & $\Gamma_{nP}^{qD}$ & $\Gamma_{nP}^{sD}$  \\ \hline
$2$	& $22(24)$ & $17(20)$	& b.t.	  & $5(4)$  	& b.t. \\
$3$	& $54(12)$ & $38(4)$	& $12(6)$	& $3(1)$	 & $1(1)$ \\
$4$	& $41(6)$	 & $25(3)$	& $14(1)$	& $1(1)$	 & $1.2(4)$ \\
$5$	& $27(4)$	 & $16(2)$	& $10(1)$	& $0.3(4)$ & $0.5(4)$ \\
$6$	& $18(3)$	 & $10(2)$	& $7(1)$	& $0.4(4)$ & $0.2(2)$ \\\hline\hline
\end{tabular}
\caption{Widths of $P$-wave charmonium states. All entries in MeV units. b.t. stands for below threshold.}
\label{wdts_cc_P}
\end{minipage}
\hfill
\begin{minipage}[t]{0.32\linewidth}
\begin{tabular}{cccc} \hline\hline
$n$ & $\Gamma^{\rm total}_{nD}$ & $\Gamma_{nD}^{qP}$ & $\Gamma_{nD}^{qP}$ \\ \hline
$2$	& $36(14)$ & $32(9)$	& $4(5)$    \\
$3$	& $42(5)$	 & $29(2)$	& $13(3)$   \\
$4$	& $34(3)$	 & $21(2)$	& $12.5(5)$ \\
$5$	& $24(3)$	 & $15(2)$	& $9(1)$    \\
$6$	& $17(3)$	 & $10(2)$	& $7(1)$    \\ \hline\hline
\end{tabular}
\caption{Widths of $D$-wave charmonium states. All entries in MeV units.}
\label{wdts_cc_D}
\end{minipage}
\end{table}

\section{Threshold EFT}\label{s:teft}

Let us consider an EFT for a quarkonium state $\psi_{nl}$ close to a heavy meson-antimeson pair threshold as the one considered in Refs.~\cite{Fleming:2008yn,Guo:2009wr,Guo:2010ak,Mehen:2011yh,Mehen:2011tp,Mehen:2013mva,Guo:2013zbw,Mehen:2015efa,Chen:2016mjn}. The heavy meson fields will be represented by $H_{\kappa}$ with $\kappa$ the spin of the light-quark state. The fields $H_{\kappa}$ carry two spin indices, the first one corresponding to the antiquark and the second one to the quark with the order being reversed for the Hermitian conjugates. Therefore, one should read expressions as ${\rm Tr}\left[H_{(1/2)}\bm{\sigma}\overline{H}_{(1/2)}\right]=(H_{(1/2)})\indices{^\beta_{\alpha_1}}\bm{\sigma}\indices{^{\alpha_1}_{\alpha_2}}(\overline{H}_{(1/2)})\indices{^{\alpha_2}_\beta}$ where $\alpha$ and $\beta$ indices correspond to the light-quark and heavy quark spin, respectively. The spin indices will be in the spherical basis unless stated otherwise. Since we work in the nonrelativistic regime, we treat the antiparticle fields (denoted with a bar) as independent fields from the particle fields. The bilinear terms in the EFT read as
\begin{align}
{\cal L}^{\rm tEFT}=\psi^\dagger_{nl}(i\pa_0-{\cal E}_{nl})\psi_{nl}+{\rm Tr}\left[H_{\kappa_1}^\dag(i\pa_0-\Lambda_{\kappa_1})H_{\kappa_1}\right]+{\rm Tr}\left[\overline{H}_{\kappa_2}^\dag(i\pa_0-\Lambda_{\kappa_2})\overline{H}_{\kappa_2}\right]\,.\label{s:teft:e5}
\end{align}

The quarkonium-heavy meson pair couplings at leading order in the heavy quark mass expansion have the following general from
\begin{align}
{\cal L}^{\rm tEFT}_{nl;\kappa^{pc}}&=\sum_{l'd}g^{(l',d)}_{nl}\left\{(-i)^{l}{\rm Tr}\left[\psi^{\dag m_l}_{nl}{\cal C}^{lm_l}_{l'm_l'\,\kappa\alpha}{\cal C}^{\kappa_1\alpha_1}_{\kappa_2\alpha_2\,\kappa\alpha}H_{\kappa_1\alpha_1}Y_{l' m_l'}(\hat{\bm{\xi}})|\bm{\xi}|^{l'+2d}\overline{H}^{\alpha_2}_{\kappa_2}\right]+{\rm h.c.}\right\}\,,
\end{align}
with $\bm{\xi}=-i\nbg=-i(\cev{\nabla}- \vec{\nabla})$ and  $\hat{\bm{\xi}}=\bm{\xi}/|\bm{\xi}|$. To match the common choice in the literature, the quarkonium field $\psi_{nl}$ is chosen to transform  under time reversal as a spherical harmonic under complex conjugation. Therefore, a factor $i^{l}$ is needed to match the time reversal transformation of a spin $l$ field and for the whole operator to be invariant under this symmetry\footnote{Notice, that $(Y_{lm_l}(\hat{\bm{\xi}}))^*=(-1)^{l-m_l}Y_{l-m_l}(\hat{\bm{\xi}})$.}. To conserve parity only the angular momentum that fulfill $p(-1)^{l+l'}=1$ are allowed. Likewise requiring charge conjugation invariance leads to the constraint $c(-1)^{l+l'}=(-1)^{l+l'+\kappa}=1$.

The couplings of quarkonium with $l=S,P,D$ to the lowest lying heavy meson-antimeson pairs ($\kappa^{p_1}_1=\kappa^{p_2}_2=(1/2)^+$ and $\kappa^{pc}=1^{--}$) up to two derivatives read as
\begin{align}
{\cal L}^{\rm tEFT}_{nS;(1^{--})}&=\frac{ig^{(P,0)}_{nS}}{\sqrt{12\pi}}{\rm Tr}\left[\psi^{\dag}_{n S}H \bm{\sigma}\cdot \nbg\overline{H}\right]\,,\label{s:teft:e1}\\
{\cal L}^{\rm tEFT}_{nP;(1^{--})}&=i\frac{g^{(S,0)}_{nP}}{\sqrt{12\pi}}{\rm Tr}\left[\psi^{\dag}_{nP}\cdot (H\bm{\sigma}\overline{H})\right]+i\frac{g^{(S,1)}_{n1}}{\sqrt{12\pi}}{\rm Tr}\left[\psi^{\dag}_{n1}\cdot (H\bm{\sigma}\nbg^2\overline{H})\right]\nn\\
&+ig^{(D,0)}_{n1}\sqrt{\frac{3}{8\pi}}{\rm Tr}\left[\psi^{\dag i}_{n1} H\bm{\sigma}^j\left(\nbg^i\nbg^j-\frac{\delta^{ij}}{3}\nbg^2\right)\overline{H}\right]\,,\label{s:teft:e2}\\
{\cal L}^{\rm tEFT}_{nD;(1^{--})}&=i\frac{g^{(P,0)}_{nD}}{\sqrt{4\pi}}{\rm Tr}\left[\psi^{\dag ij}_{nD}(H \bm{\sigma}^{j}\nbg^{i}\overline{H})\right]\,.\label{s:teft:e3}
\end{align}
The spin indices in Eqs.~\eqref{s:teft:e1}-\eqref{s:teft:e3}, explicit or implicit, are in the Cartesian basis. The first two terms can be found in Refs.~\cite{Fleming:2008yn,Guo:2009wr,Guo:2010ak,Mehen:2011yh} with different normalizations for the couplings. The equivalence with Refs.~\cite{Fleming:2008yn,Mehen:2011yh} is
\begin{align}
g_1=\frac{g^{(S,0)}_{11}}{\sqrt{3\pi}}\,,\quad g_2=\frac{g^{(P,0)}_{10}}{\sqrt{3\pi}}\,,
\end{align}
for Refs.~\cite{Guo:2009wr,Guo:2010ak} an extra minus sign is needed to account for a different definition of $\nbg$. 

Now we match the threshold EFT to BOEFT. At the tree level the matching can be obtained expanding the fields in eigenstates of the relative motion Hamiltonian. For the quarkonium field for instance
\begin{align}
\Psi(t,\,\bm{R},\,\bm{r})=\sum_{nl}(\Psi_{nl}(t,\,\bm{R}))^{m_l}(\psi_{nl}(\bm{r}))_{m_l}\,,\label{mtt:e5}
\end{align}
with $\psi_{nl}$ defined in Eq.~\eqref{s:ths:e4}. In the same way we expand the heavy meson-antimeson pair field. Since in this case the eigenfunctions are plane waves we have
\begin{align}
{\cal M}_{\kappa\alpha}(t,\,\bm{R},\,\bm{r})=\int\frac{d^3k}{(2\pi)^3}({\cal M}_{\kappa\bm{k}}(t,\,\bm{R}))_{\alpha}e^{-i\bm{k}\cdot\bm{r}}\,.\label{mtt:e1}
\end{align}
Introducing the partial wave expansion of Eq.~\eqref{mtt:e2} to Eq.~\eqref{mtt:e1} and after some manipulations we obtain
\begin{align}
{\cal M}_{\kappa\alpha}(t,\,\bm{R},\,\bm{r})&=\sum_{\ell}\sum^{\ell+\kappa}_{l=|\ell-\kappa|}\int\frac{d^3k}{(2\pi)^3}\left(({\cal M}_{\kappa\bm{k}}(t,\,\bm{R}))^{\alpha'}Y^{m_l'}_{l}(\hat{\bm{k}})\right){\cal C}^{\ell m_\ell}_{lm_l'\,\kappa\alpha'}{\cal C}^{\ell m_\ell}_{lm_l\,\kappa-\alpha}\left((-1)^{\kappa-\alpha}4\pi i^{-l}j_l(kr)Y_{lm_l}(\hat{\bm{r}})\right)\,.\label{mtt:e3}
\end{align}
So far we have made no approximation. Now, we note that in the case a quarkonium state mass is close to that of a heavy meson-antimeson pair then the relative momentum between the heavy quarks in the quarkonium state $\sim 1/r$ is larger than the relative momentum between the heavy mesons $k$. Therefore, we can expand the spherical Bessel function for $kr\ll 1$:
\begin{align}
4\pi j_l(kr)&=\sum_d b^d_l (kr)^{2d+l}\,,\\
b^d_l&=\frac{4\pi(-1)^d2^l(d+l)!}{d!(2d+2l+1)!}\,.
\end{align}
Using this expansion in Eq.~\eqref{mtt:e3} all the dependence on $r$ factorizes. After a few more manipulations we arrive at
\begin{align}
\eqref{mtt:e3}=\sum_{\ell}\sum^{\ell+\kappa}_{l=|\ell-\kappa|}\sum_d\left({\cal C}^{\ell m_\ell}_{lm_l\,\kappa-\alpha}(-1)^{\kappa-\alpha}b^d_{l}r^{l+2d}Y_{lm_l}(\hat{\bm{r}})\right)\left((-1)^{\ell+m_\ell}\int\frac{d^3k}{(2\pi)^3}{\cal C}^{\ell -m_\ell}_{lm_l'\,\kappa\alpha'}({\cal M}_{\kappa\bm{k}}(t,\,\bm{R}))_{\alpha'}Y_{lm_l'}(\hat{\bm{k}})|\bm{k}|^{l+2d}\right)\,.
\end{align}
We set the following matching condition
\begin{align}
\int\frac{d^3k}{(2\pi)^3}\left(({\cal M}_{\kappa\bm{k}})_{\alpha'}(t,\,\bm{R})Y_{lm_l'}(\hat{\bm{k}})|\bm{k}|^{l+2d}\right)\cong{\cal C}^{\kappa_1\alpha_1}_{\kappa_2\alpha_2\,\kappa\alpha'}H_{\kappa_1\alpha_1}(t,\,\bm{R})Y_{l m_l'}(\hat{\bm{\xi}})|\bm{\xi}|^{l+2d}\overline{H}^{\alpha_2}_{\kappa_2}(t,\,\bm{R})\,.
\end{align}

\begin{figure}[ht!]
\centerline{\includegraphics[width=0.7\linewidth]{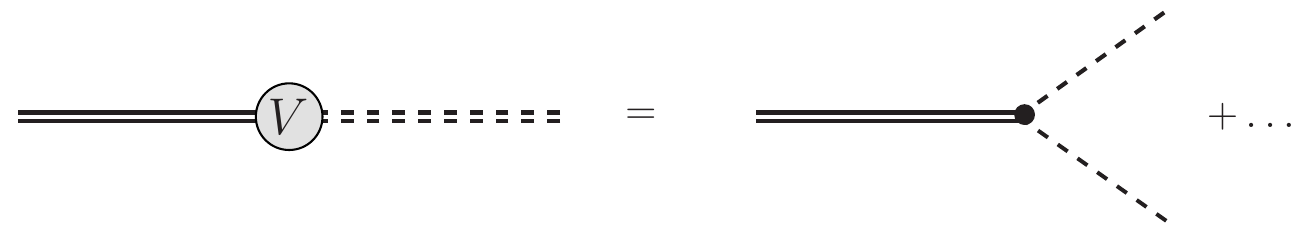}}
\caption{Matching of the quarkonium-heavy meson pair vertex between BOEFT (left) and the threshold EFT (right). The double continuous lines represent the quarkonium state, the double dashed line represents the heavy meson pair field in BOEFT and the single dashed lines correspond to the heavy mesons in the threshold EFT. The dots in the right-hand side stand for the vertices with extra derivatives.}
\label{tree_lvl_match}
\end{figure}

We can now apply these field expansions into the BOEFT couplings and obtain the tree level matching of the couplings of quarkonium to heavy meson pairs in the threshold EFT. The matching is shown diagrammatically in Fig.~\ref{tree_lvl_match}. We arrive at 
\begin{align}
&\int dr^3\Psi^\dag V^{\kappa}_{\rm mix}(P^*_{\kappa 0})^\alpha{\cal M}_{\kappa\alpha}\cong \sum_{l}\sum^{l+\kappa}_{l'=|l-\kappa|}\sum_d g^{(l',d)}_{nl}(-i)^l\psi^{\dag m_l}_{nl}
{\cal C}^{lm_l}_{l'm_l'\,\kappa\alpha}{\cal C}^{\kappa_1\alpha_1}_{\kappa_2\alpha_2\,\kappa\alpha}H_{\kappa_1\alpha_1}Y_{l' m_l'}(\hat{\bm{\xi}})|\bm{\xi}|^{l'+2d}\overline{H}^{\alpha_2}_{\kappa_2}\,,
\end{align}
with
\begin{align}
g^{(l',d)}_{nl}={\cal C}^{l 0}_{l'0\,\kappa 0}\sqrt{\frac{(2l'+1)}{(2l+1)}}b^d_{l'}\int dr r^{2+l'+2d}\phi_{n l}(r) V^{\kappa}_{\rm mix}(r)\,.\label{s:teft:e4}
\end{align}

\begin{figure}[ht!]
\centerline{\includegraphics[width=0.9\linewidth]{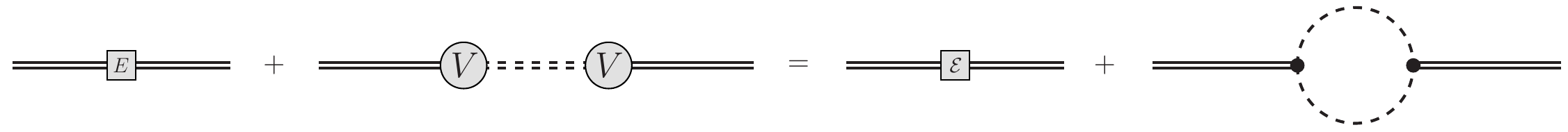}}
\caption{Matching of the quarkonium two-point function between BOEFT (left) and the threshold EFT (right). The double continuous lines represent the quarkonium, the double dashed line represent the heavy meson pair field in BOEFT while the single dashed lines correspond to the heavy mesons in the threshold EFT.}
\label{one_loop_match}
\end{figure}

It is interesting to consider also the matching of the quarkonium bilinear. At tree level these can be easily obtained using the expansion in Eq.~\eqref{mtt:e5}. However, in this case one should also consider the self-energy contribution which appears at one loop. The matching is shown diagrammatically in Fig.~\ref{one_loop_match}. The key point is to identify the momentum region in the loop diagram in the BOEFT side (left) that matches the heavy meson loop in the threshold EFT side (right). The BOEFT diagram is given in Eq.~\eqref{mw:e3} and contains two momentum scales $k\sim k_d$ and $k\sim 1/r$. The contribution of the first one matches the heavy meson loop, while the latter gives a contribution to the residual mass of the quarkonium state in the Lagrangian in Eq.~\eqref{s:teft:e5}
\begin{align}
{\cal E}_{nl}=E_{nl}+\sum^{l+1}_{l'=|l-1|}{\cal E}^{\kappa l'}_{nl}\,,
\end{align}
where $E_{nl}$ is the energy of the quarkonium state computed in BOEFT including the heavy meson-antimeson pair contributions except the one of the nearby heavy meson-antimeson pair considered explicitly in the threshold EFT and
\begin{align}
{\cal E}^{\kappa l'}_{nl}=\frac{4\mu}{\pi}\int^{\infty}_0 dk\, [a^{\kappa l'}_{nl}(k)]^2\,.\label{mtt:e6}
\end{align}

To determine if the threshold expansion can be carried out, we compare the values of the $|\bm{k}_d|$ momentum scale of the mesons with the value of $\langle 1 /r \rangle$ for the quarkonium states in Table~\ref{c_to_t}. The computation of $|\bm{k}_d|$ is carried out for the transitions between quarkonium and the lowest lying heavy meson pair thresholds without and with closed strangeness. The value of $\langle 1 /r \rangle$ is computed including ${\cal O}(1/m_Q)$ corrections to the quarkonium wave function and considering intermediate states up to $n=8$. In principle, these corrections could be sizable for charmonium states since $\Lambda_{\rm QCD}/m_c\sim 18\%$. However, we find that only the ground states get contributions comparable to this parametric estimate. In practice, for excited states the contributions from lower lying intermediate states almost cancel with those of higher laying states leading to very small contributions. Therefore, we expect the parametric estimate of ${\cal O}(1/m^2_Q)$ contributions to $\langle 1 /r \rangle$, i.e. $(\Lambda_{\rm QCD}/m_c)\sim 3.5\%$ and  $(\Lambda_{\rm QCD}/m_b)\sim 0.4\%$, to be significantly larger than actual contributions. The computation of $|\bm{k}_d|$ on the other hand involves uncertainties inherited from the uncertainty in the determination of the quarkonium masses. Unfortunately, the effect is naturally more significant for states close to threshold. Therefore, at the present accuracy in the determination of the quarkonium masses, we can only rule out the threshold expansion validity for certain states while in no case it can be completely confirmed as a good approximation. For the latter cases we compute the values of the quarkonium-heavy-meson-pair couplings, which we collect in Table~\ref{teft_cpl}. The value of these couplings only depend on the quarkonium bound state through the wave function, and therefore we expect our values to be reliable despite the uncertainty on the quarkonium masses. Nevertheless, due to possible ${\cal O}(1/m_Q)$ contributions to the mixing potential, the accuracy of the couplings is limited to corrections ${\cal O}(\Lambda_{\rm QCD}/m_Q)$, which is reflected in the uncertainty. Next, we compute the contribution to the residual mass of the quarkonium field in the threshold EFT, ${\cal E}^{\kappa l'}_{nl}$ defined in Eq.~\eqref{mtt:e6}. This quantity also only depends on the quarkonium state through the wave function and its uncertainty is assessed as of the size of corrections ${\cal O}(\Lambda_{\rm QCD}/m_Q)$. The values of ${\cal E}^{\kappa l'}_{nl}$ can be found in Table~\ref{teft_cpl} for bottomonium and charmonium.

\begin{table}[ht!]
\begin{minipage}[t]{0.49\linewidth}
\begin{tabular}{ccccc} \hline\hline
$l$ & $n$ & $\langle 1 /r \rangle$ & $|\bm{k}_d|(B\bar{B})$ & $|\bm{k}_d|(B_s\bar{B}_s)$ \\ \hline
$0$ &	$1$ &	$1.311$ &	$2.389^{+87}_{-91}$   &	$2.604^{+82}_{-84}$   \\
$0$ &	$2$ &	$0.672$ &	$1.768^{+116}_{-125}$ &	$2.038^{+103}_{-109}$ \\
$0$ &	$3$ &	$0.503$ &	$1.168^{+170}_{-199}$ &	$1.536^{+135}_{-148}$ \\
$0$ &	$4$ &	$0.408$ &	$0.353^{+389}_{-353}$ &	$0.920^{+211}_{-277}$ \\
$0$ &	$5$ &	$0.343$ &	$1.210^{+165}_{-191}$ &	$0.718^{+255}_{-429}$ \\
$0$ &	$6$ &	$0.292$ &	$1.642^{+125}_{-135}$ &	$1.330^{+154}_{-174}$ \\
$1$ &	$1$ &	$0.589$ &	$1.916^{+108}_{-114}$ &	$2.169^{+97}_{-102}$  \\
$1$ &	$2$ &	$0.447$ &	$1.371^{+147}_{-165}$ &	$1.698^{+123}_{-132}$ \\
$1$ &	$3$ &	$0.372$ &	$0.599^{+286}_{-599}$ &	$1.157^{+174}_{-205}$ \\
$1$ &	$4$ &	$0.321$ &	$1.001^{+194}_{-242}$ &	$0.214^{+478}_{-214}$ \\
$1$ &	$5$ &	$0.280$ &	$1.502^{+135}_{-149}$ &	$1.149^{+175}_{-207}$ \\
$1$ &	$6$ &	$0.251$ &	$1.850^{+111}_{-119}$ &	$1.584^{+131}_{-143}$ \\
$2$ &	$1$ &	$0.415$ &	$1.578^{+129}_{-141}$ &	$1.872^{+112}_{-119}$ \\
$2$ &	$2$ &	$0.348$ &	$0.960^{+200}_{-255}$ &	$1.382^{+148}_{-166}$ \\
$2$ &	$3$ &	$0.304$ &	$0.689^{+260}_{-466}$ &	$0.699^{+260}_{-461}$ \\
$2$ &	$4$ &	$0.271$ &	$1.323^{+152}_{-172}$ &	$0.899^{+215}_{-286}$ \\
$2$ &	$5$ &	$0.244$ &	$1.716^{+120}_{-129}$ &	$1.422^{+145}_{-161}$ \\
$2$ &	$6$ &	$0.219$ &	$2.018^{+103}_{-108}$ &	$1.780^{+118}_{-126}$ \\ \hline\hline
\end{tabular}
\end{minipage}
\hfill
\begin{minipage}[t]{0.49\linewidth}
\begin{tabular}{ccccc} \hline\hline
$l$ & $n$ & $\langle 1 /r \rangle$ & $|\bm{k}_d|(D\bar{D})$ & $|\bm{k}_d|(D_s\bar{D}_s)$ \\ \hline
$0$ &	$1$ &	$0.724$ &	$1.254^{+61}_{-65}$   &	$1.443^{+56}_{-59}$   \\
$0$ &	$2$ &	$0.447$ &	$0.624^{+116}_{-143}$ &	$0.915^{+87}_{-96}$   \\
$0$ &	$3$ &	$0.347$ &	$0.725^{+102}_{-119}$ &	$0.354^{+186}_{-354}$ \\
$0$ &	$4$ &	$0.290$ &	$1.149^{+67}_{-71}$   &	$0.981^{+81}_{-89}$   \\
$0$ &	$5$ &	$0.250$ &	$1.429^{+54}_{-56}$   &	$1.312^{+62}_{-65}$   \\
$0$ &	$6$ &	$0.218$ &	$1.645^{+47}_{-49}$   &	$1.556^{+53}_{-54}$   \\
$1$ &	$1$ &	$0.380$ &	$0.870^{+86}_{-96}$   &	$1.106^{+73}_{-78}$   \\
$1$ &	$2$ &	$0.300$ &	$0.445^{+151}_{-244}$ &	$0.468^{+153}_{-238}$ \\
$1$ &	$3$ &	$0.255$ &	$1.010^{+75}_{-81}$   &	$0.803^{+98}_{-111}$  \\
$1$ &	$4$ &	$0.225$ &	$1.326^{+58}_{-61}$   &	$1.192^{+68}_{-72}$   \\
$1$ &	$5$ &	$0.201$ &	$1.561^{+50}_{-51}$   &	$1.462^{+56}_{-58}$   \\
$1$ &	$6$ &	$0.183$ &	$1.754^{+44}_{-46}$   &	$1.676^{+49}_{-50}$   \\
$2$ &	$1$ &	$0.279$ &	$0.413^{+160}_{-300}$ &	$0.779^{+100}_{-115}$ \\
$2$ &	$2$ &	$0.237$ &	$0.827^{+90}_{-102}$  &	$0.540^{+137}_{-186}$ \\
$2$ &	$3$ &	$0.210$ &	$1.202^{+64}_{-68}$   &	$1.046^{+77}_{-83}$   \\
$2$ &	$4$ &	$0.190$ &	$1.465^{+53}_{-55}$   &	$1.353^{+60}_{-63}$   \\
$2$ &	$5$ &	$0.174$ &	$1.672^{+47}_{-48}$   &	$1.586^{+52}_{-53}$   \\
$2$ &	$6$ &	$0.161$ &	$1.847^{+42}_{-43}$   &	$1.779^{+46}_{-47}$   \\ \hline\hline
\end{tabular}
\end{minipage}
\caption{The expected value $\langle 1/r\rangle$ and the on-shell momentum $|\bm{k}_d|$ for charmonium (left) and bottomonium (right) states with angular and principal quantum numbers $l$ and $n$. The value $|\bm{k}_d|$ corresponds to the momentum of the heavy mesons in the transition from a quarkonium state to the heavy meson pair. The expected value $\langle 1/r\rangle$ is computed up to contributions of ${\cal O}(1/m_Q)$ to the wave function. The superindex and subindex of the $|\bm{k}_d|$ values indicate the difference between the maximum and minimum values, respectively, within the uncertainty of the quarkonium state mass. The threshold expansion is only valid when $|\bm{k}_d|<<\langle 1/r\rangle$. All quantities are in GeV units.}
\label{c_to_t}
\end{table}

\begin{table}[ht!]
\begin{minipage}[t]{0.49\linewidth}
\begin{tabular}{ccccccc} \hline\hline
$l$  & $n$ & $M\overline{M}$ & $l'$ &  $g^{(l',0)}_{nl}$ & $g^{(l',1)}_{nl}$ & ${\cal E}^{l'}_{nl}$~[MeV] \\ \hline
$0$  & $4$ & $B\bar{B}$     &	$P$  &	$-4.2(3)$    &	$15(1)$     &	$-27(2)$         \\
$0$  & $5$ & $B_s\bar{B}_s$ &	$P$  &	$3.6(2)$     &	$-21(1)$    &	$-8.2(5)$        \\
$1$  & $3$ & $B\bar{B}$     &	$S$  &	$-1.4(7)$    &	$7.1(4)$   &	$-35(2)$         \\
$1$  & $3$ & $B\bar{B}$     &	$D$  &	$4.0(2)$     &	$-10.1(6)$  &	$-17(1)$         \\
$1$  & $4$ & $B_s\bar{B}_s$ &	$S$  &	$0.82(5)$    &	$-7.9(5)$   &	$-7.2(4)$        \\
$1$  & $4$ & $B_s\bar{B}_s$ &	$D$  &	$-4.5(3)$    &	$16(1)$     &	$-5.2(3)$        \\
$2$  & $3$ & $B\bar{B}$     &	$P$  &	$-2.7(1)$    &	$10.9(6)$   &	$-17(1)$         \\
$2$  & $3$ & $B_s\bar{B}_s$ &	$P$  &	$-2.7(2)$    &	$10.9(7)$   &	$-8.5(5)$        \\ \hline\hline
\end{tabular}
\end{minipage}
\hfill
\begin{minipage}[t]{0.49\linewidth}
\begin{tabular}{ccccccc} \hline\hline
$l$  & $n$ & $M\overline{M}$     & $l'$ &  $g^{(l',0)}_{nl}$ & $g^{(l',1)}_{nl}$ & ${\cal E}^{l'}_{nl}$~[MeV] \\ \hline
$0$  & $3$ & $D_s\bar{D}_s$ &	$P$  &	$4.5(9)$     &	$-29(5)$    &	$-6(1)$          \\
$1$  & $2$ & $D\bar{D}$     &	$S$  &	$0.4(1)$     &	$-11(2)$    &	$-17(3)$         \\
$1$  & $2$ & $D\bar{D}$     &	$D$  &	$-6(1)$      &	$19(4)$     &	$-10(2)$         \\
$2$  & $1$ & $D\bar{D}$     &	$P$  &	$-5(1)$      &	$11(2)$     &	$-23(4)$         \\ \hline\hline
\end{tabular}
\end{minipage}
\caption{Couplings $g^{(l',d)}$ of the charmonium (left) and bottomonium (right) states with quantum numbers $(l,n)$ to heavy mesons in the $l'$ partial wave. We display only the cases for which $|\bm{k}_d|< \langle 1 /r \rangle$ within uncertainty as displayed in Table~\ref{c_to_t}. All dimension-full entries are in GeV unless indicated otherwise. Note that the couplings for $B\bar{B}$ and $D\bar{D}$ pairs apply to both neutral and charged meson pair cases.}
\label{teft_cpl}
\end{table}

\section{Conclusions}\label{sec:con}

The first excited state in the spectrum of static energies in the quarkonium sector corresponds to a heavy meson-antimeson pair~\cite{Bali:2005fu,Bulava:2019iut}. Pairs composed of heavier heavy mesons appear as subsequent excited states, along with quarkonium hybrid static energies. The heavy meson-antimeson static energies are characterized by the total spin of the light quarks and its projection onto the heavy quark-antiquark axis, thus each heavy meson-antimeson pair is associated with more than one static state. EFTs have been built to describe the heavy quark-antiquark bound states supported by the spectrum of static energies. These EFTs incorporate the heavy quark mass expansion and the adiabatic expansion between the heavy and light degrees of freedom. For standard quarkonium the EFT is known as strongly coupled potential NRQCD~\cite{Brambilla:2000gk,Pineda:2000sz}, and its extension to nontrivial light degrees of freedom is called Born-Oppenheimer EFT (BOEFT)~\cite{Berwein:2015vca,Oncala:2017hop,Brambilla:2017uyf,Soto:2020pfa,Soto:2020xpm}.

In this paper we have shown how to incorporate into BOEFT the heavy meson-antimeson pairs and have obtained all the leading order operators coupling them to quarkonium. The Lagrangian containing these couplings can be found in Eq.~\eqref{sec:ths:e2}. Using these couplings, we have obtained the expressions for the contribution of heavy meson-antimeson pairs to the quarkonium masses and decay widths in perturbation theory. These formulas, in Eqs.~\eqref{mw:e4}, \eqref{mw:e5} and \eqref{mw:e6}, depend on the total spin of the light-quarks in the threshold state that couples to quarkonium, the mixing potential accompanying the coupling operator, the mass gap between the threshold and the quarkonium state and the wave function of the quarkonium state.

In Sec.~\ref{sec:mnrq} we have discussed the matching of the new potentials, in particular of the mixing one, to NRQCD. The matching has been obtained for both when the mixing potentials can be considered a perturbation and when not. We have also shown that the second case reduces to the first when the separation between the static potentials of the quarkonium and the heavy meson-antimeson pair is larger than the mixing potential, which is the case for most of the range of $r$ except for a small region around the string breaking distance. In Ref.~\cite{Bali:2005fu} the ground and first excited states for the coupled system of quarkonium with the lowest lying heavy heavy meson-antimeson pair were obtained in lattice QCD. Using this data and Eqs.~\eqref{s:match:e1}-\eqref{s:match:e3}, the quarkonium and heavy meson pair static potentials as well as the mixing potential can be obtained. It is interesting that the small bump in the excited state at short distances, see Fig.~\ref{lat_data}, which could be interpreted as a heavy meson-antimeson interaction, disappears completely in the heavy meson-antimeson static potential in Fig.~\ref{QQbpotplot}. This highlights the importance of taking into account the mixing with heavy quark-(anti)quark states when studying the heavy meson-(anti)meson interactions.

We computed the contribution of the lowest lying heavy meson-antimeson pairs without and with closed strangeness to the masses and widths of the bottomonium and charmonium states with $l=S,\,P,\,D$ and $n=1,..,6$ covering the mass range where exotic quarkonium states have been discovered. The quarkonium static potential is obtained combining fix order results in the RS' scheme for the short-distance part of the potential and a fit to lattice data for the medium and long-ranges. To increase the accuracy in the determination of the quarkonium spectrum in the threshold region we compute the quarkonium spectra up to ${\cal O}(1/m_Q)$. The quarkonium $1/m_Q$ suppressed potential is also parametrized using perturbation theory for the short-distance part and a fit to lattice data~\cite{Koma:2006si,Koma:2007jq} for the remaining range of $r$. Our results for the quarkonium masses and the contribution of the lowest lying thresholds are shown in Tables~\ref{sp_bb_s}-\ref{sp_bb_d} for bottomonium and Tables~\ref{sp_cc_s}-\ref{sp_cc_d} for charmonium. Our result show the contribution of these two thresholds to the quarkonium masses is comparable to that of the $1/m_Q$ suppressed potential. The uncertainty associated to the heavy quark mass can be eliminated by shifting the bottomonium and charmonium spectra to match the experimental mass of a given reference state at the price of not giving a prediction for this reference state.  In Table~\ref{sp_sft} we show such spectra taking as a reference the spin-average of the $2S$ doublet. The resulting spectra is compared to experimental values in Figs.~\ref{plt_bbbar_spc} and \ref{plt_ccbar_spc}. For the widths we find values of about $5-10$~MeV for bottomonium, in Tables~\ref{wdts_bb_S}-\ref{wdts_bb_D}, and $10-50$~MeV, in Tables~\ref{wdts_cc_S}-\ref{wdts_cc_D}, for charmonium. 

Unfortunately, the contributions of the thresholds to a quarkonium state mass and width become very imprecise when the mass gap between them is similar to the uncertainty in the quarkonium mass determination. To improve the accuracy of the threshold contributions, the heavy-quark spin dependent contributions should be taken into account. These can be split between the ones affecting the heavy meson masses, which are of ${\cal O}(1/m_Q)$, and the ones to the quarkonium masses which are ${\cal O}(1/m^2_Q)$. For an accuracy of a few MeV the former ones should be enough in the bottomonium sector but both would be necessary for charmonium states.

In Sec.~\ref{s:teft}, we have discussed the matching of BOEFT with heavy meson-antimeson degrees of freedom to a threshold EFT containing as explicit degrees of freedom a quarkonium state and the heavy mesons of a nearby threshold interacting through contact operators. This is possible when the relative momentum of the heavy mesons is smaller than the inverse of the size of the quarkonium state. The latter being a measure of the relative momentum between the heavy quarks in the quarkonium state. We obtained the matching expression, in terms of the mixing potential and the quarkonium wave function, of the series of couplings with increasing derivatives of a quarkonium state to the heavy meson-antimeson pair, which can be found in Eq.~\eqref{s:teft:e4}. At the current level of accuracy we are only able to rule out for which states and heavy meson thresholds this matching is not valid. Nevertheless, we provide the values of the couplings to the two lowest lying meson-antimeson pairs with the quarkonium states which are not ruled out in Table~\ref{teft_cpl}. This is possible since the values of the couplings do not depend directly on the mass gap between the threshold and the quarkonium state.

Using the threshold EFT and following the analysis from Refs.~\cite{Braaten:2003he,Soto:2009xy,Matuschek:2020gqe,Habashi:2020ofb} one can study the heavy meson molecule picture for exotic states close to a heavy meson-antimeson threshold. As pointed out in Ref.~\cite{Braaten:2003he} the molecular nature of $X(3872)$ can be explained by an accidental fine-tuning of $\chi_{1}(2P)$ mass to the $\bar{D}D^*$ threshold which would result in a abnormally large scattering length for the heavy meson-antimeson scattering. Such scenario is compatible with our results, however to confirm it, it would require a high precision determination of the $\chi_{1}(2P)$ mass.

Hybrid quarkonium states also are expected to appear in the threshold region. The hybrid states associated to the lowest lying gluelump, with $\kappa^{pc}=1^{+-}$ , appear at $4.000-4.150$~GeV and $10.690-10.790$~GeV~\cite{Berwein:2015vca,Oncala:2017hop,Brambilla:2019jfi} in the charmonium and bottomonium sectors, respectively. Meanwhile, the ones associated to the second lowest lying gluelump, with $\kappa^{pc}=1^{--}$, appear at $4.5$~GeV and $11.14$~GeV~\cite{Pineda:2019mhw} in the charmonium and bottomonium sectors, respectively. For the former case the mixing with quarkonium is heavy quark mass suppressed~\cite{Oncala:2017hop} and its effects are of ${\cal O}(1/m^2_Q)$. However, in the latter case the mixing is not suppressed and its effects can be, in principle, of the same order as the heavy meson-antimeson pair of similar mass. Therefore, its study would be interesting. Furthermore, for an accurate computation of their hybrid quarkonium masses, which is also necessary for a good understanding of the spectrum of quarkonium-like states in the threshold region, it is also necessary the study of the mixing of quarkonium hybrids with the heavy meson-antimeson pairs. For this objective, the formulation of the mixing terms of hybrid quarkonium states with heavy meson-antimeson pairs should be worked out as well as the matching expression as NRQCD correlators. The latter should be computed with lattice QCD or models in order to obtain numerical evaluations.

\section*{Acknowledgments}                             

J.T.C. thanks Joan Soto for providing the lattice data from Refs.~\cite{Koma:2006si,Koma:2007jq,Koma:2012bc,Bulava:2019iut} and acknowledges financial support by National Science Foundation (No. PHY-2013184). 

\appendix

\section{Spin notation and time reversal}\label{a1}

Let $(a_\kappa)_\alpha$ be the covariant components of an irreducible tensor of rank $\kappa$ and $(b_\kappa)^{\alpha}$ the contravariant components of an irreducible tensors of rank $\kappa$. Under rotations, the latter transforms with the complex conjugate transformation of the former. To differentiate between the two transformations we write the spin indices that transform with the complex conjugate representation as superindices and the ones that transform as the standard representation as lower indices.
\begin{align}
(a_\kappa)_\alpha&=(D_{\kappa})\indices{_\alpha^{\alpha'}}(a_\kappa)_{\alpha'}\,,\\
(b_\kappa)^\alpha&=(D^*_{\kappa})\indices{^\alpha_{\alpha'}}(b_\kappa)^{\alpha'}\,,
\end{align}
with $(D_{\kappa})\indices{_\alpha^{\alpha'}}$ a Wigner function. Recall that repeated spin indices should be understood as summed.

We follow standard conventions and write irreducible tensor fields for spin $\kappa$ particles and antiparticles in covariant and contravariant basis, respectively. However, it should be noted that a covariant basis for both is also possible. For instance, for the heavy antiquark these two choices correspond to using $\chi$ or $\chi_c=i\sigma_2\chi^*$ as our heavy antiquark field. Similarly, the light-quark operators $\overline{\cal Q}_{\kappa^{p}}$ could be replaced by the charge conjugates of ${\cal Q}_{\kappa^{p}}$ and this way both light-quark operators would transform in the covariant basis.

Sums over the same spin index as superscript and subscript are invariant under rotation transformations
\begin{align}
(b_\kappa)^\alpha(a_\kappa)_\alpha=(b_\kappa)^{\alpha'}(D^\dag_{\kappa})\indices{_{\alpha'}^\alpha}(D_{\kappa})\indices{_\alpha^{\alpha''}}(a_\kappa)_{\alpha''}=(b_\kappa)^{\alpha'}\delta\indices{_{\alpha'}^{\alpha''}}(a_\kappa)_{\alpha''}=(b_\kappa)^{\alpha'}(a_\kappa)_{\alpha'}\,.
\end{align}
We can lower or rise an index by applying the transformation~\cite{fano:1959}
\begin{align}
(a_\kappa)^\alpha&=(\aleph_\kappa)_{\alpha\alpha'}(a_\kappa)_{\alpha'}=(-1)^{\kappa-\alpha}(a_\kappa)_{-\alpha}\,,\\
(b_\kappa)_\alpha&=(\aleph_\kappa)_{\alpha\alpha'}(b_\kappa)^{\alpha'}=(-1)^{\kappa-\alpha}(b_\kappa)^{-\alpha}\,,
\end{align}
with 
\begin{align}
(\aleph_\kappa)_{\alpha\alpha'}=D^{\kappa}_{\alpha\alpha'}(0,\pi,0)=e^{-i\pi (S^y_{\kappa})_{\alpha\alpha'}}\,.
\end{align}
If the components of an irreducible tensor are complex numbers then the $\aleph_\kappa$ transformation is equivalent to complex conjugation.

The scalar product is defined as
\begin{align}
(a_\kappa)\cdot (a_\kappa')\equiv (a_\kappa)^\alpha (a_\kappa')_\alpha=(-1)^{\kappa-\alpha}(a_\kappa)_{-\alpha}(a_\kappa')_\alpha\,.
\end{align}
The irreducible product of two covariant irreducible tensors is
\begin{align}
\{a_{\kappa_1}\otimes a_{\kappa_2}\}_{\kappa\alpha}\equiv{\cal C}^{\kappa\alpha}_{\kappa_1\alpha_1\,\kappa_2\alpha_2}(a_{\kappa_1})_{\alpha_1} (a_{\kappa_2})_{\alpha_2}\,.\label{a1:e1}
\end{align}
Note that using Eq.\eqref{a1:e1} to form a scalar out of to irreducible tensors of the same rank yields a different normalization than the scalar product
\begin{align}
\{a_{\kappa}\otimes a_{\kappa}'\}_{00}=\frac{(-1)^{2\kappa}}{\sqrt{2\kappa+1}}(a_\kappa)\cdot (a_\kappa')\,.
\end{align}
To combine a covariant and a contravariant irreducible tensors into an irreducible representation one of the irreducible tensors needs to be transformed to match the transformation of the other~\cite{fano:1959}. We choose the following
\begin{align}
\{a_{\kappa_1}\otimes \aleph_{\kappa_2} b_{\kappa_2}\}_{\kappa\alpha}={\cal C}^{\kappa\alpha}_{\kappa_1\alpha_1\,\kappa_2-\alpha_2}(-1)^{\kappa_2+\alpha_2}(a_{\kappa_1})_{\alpha_1}(b_{\kappa_2})^{\alpha_2}\,.
\end{align}
However, other forms are also valid. For instance, since
\begin{align}
(-1)^{\kappa_2+\alpha_2}{\cal C}^{\kappa\alpha}_{\kappa_1\alpha_1\,\kappa_2-\alpha_2}=(-1)^{2\kappa_2}\sqrt{\frac{2\kappa+1}{2\kappa_1+1}}{\cal C}^{\kappa_1\alpha_1}_{\kappa_2\alpha_2\,\kappa\alpha}\,,
\end{align}
the product
\begin{align}
{\cal C}^{\kappa_1\alpha_1}_{\kappa_2\alpha_2\,\kappa\alpha}(a_{\kappa_1})_{\alpha_1}(b_{\kappa_2})^{\alpha_2}\,,
\end{align}
is also an irreducible tensor of rank $\kappa$.

The time reversal operator~\cite{Weinberg:1995mt} is
\begin{align}
T=(\aleph_\kappa) K\,,
\end{align}
with $K$ the complex conjugate operator. This form of the time reversal operator also relies of a standard form of the spin matrices in which $J_y$ is purely imaginary. We have chosen a $(1/2)^*\otimes(1/2)$ representation of the heavy-quark spin indices, which is convenient since we do not have to specify if the heavy quark spin state is a singlet or a triplet. The time reversal transformation on the fields is as follows
\begin{align}
T \Psi(t,\,\bm{r},\,\bm{R})T^{-1}&=\sigma_2\Psi(-t,\,\bm{r},\,\bm{R})\sigma_2\,,\\
T{\cal M}_{\kappa \alpha}(t,\,\bm{r},\,\bm{R})T^{-1}&=(-1)^{\kappa-\alpha}\sigma_2{\cal M}_{\kappa -\alpha}(-t,\,\bm{r},\,\bm{R})\sigma_2\,.
\end{align}
The spherical harmonics with the Condon-Shortley normalization transform as
\begin{align}
KY_{lm}&=Y^*_{lm}=(-1)^mY_{l-m}\,,\\
(\aleph_{l})_{mm'}Y_{lm'}&=(-1)^{l-m}Y_{l-m}\,.
\end{align}
Therefore, it is convenient to always consider $i^l Y_{lm}$ in the construction of operators in the Lagrangian. In this way scalar products such as
\begin{align}
(b_l)^m(i^lY_{lm})\,,\quad (i^lY_{lm})^*(a_l)_m\,,
\end{align}
are invariant under time reversal symmetry.

\bibliographystyle{apsrev4-2}
\bibliography{hmtbib}

\end{document}